\documentclass[ALICE,manyauthors]{cernphprep}
\usepackage{color} 
\usepackage{bm}

\usepackage{cite}
\newcommand{\jpsi}{\rm J/$\psi$}
\newcommand{\psip}{$\psi$(2S)}

\newcommand{\pt}{$p_{{\mathrm T}}$}

\newcommand{\RpPb}{$R_{\mathrm{pPb}}$}

\newcommand{\sqrts}{$\sqrt{s_{\rm NN}} = 5.02$ TeV}
\newcommand{\sqrtspp}{$\sqrt{s} = 5.02$ TeV}

\newcommand{\pPb}{$\mbox{p-Pb}$}

\newcommand{\ccbar}{$c\overline{c}$}
\usepackage{rotating}
\usepackage{lineno}
\usepackage{hyperref}

\begin{document}%
%
%
\begin{titlepage}
\PHnumber{2015-030}      
\PHdate{February 12, 2015}            
%
%
\title{Rapidity and transverse-momentum dependence of the inclusive J/$\mathbf{\psi}$ nuclear
modification factor in \mbox{p-Pb} collisions at $\mathbf{\sqrt{{\textit s}_{NN}}}=5.02$ TeV}

\ShortTitle{Rapidity and transverse-momentum dependence of the inclusive J/$\psi$ $R_{\rm pPb}$}
%
\Collaboration{ALICE Collaboration}%
\ShortAuthor{ALICE Collaboration}      
\begin{abstract}
{We have studied the transverse-momentum (\pt) dependence of the inclusive \jpsi\ production in 
\mbox{p--Pb} collisions at $\sqrt{s_{\rm NN}} = 5.02$ TeV, in three center-of-mass rapidity ($y_{\rm cms}$) regions, down to zero \pt. 
Results in the forward and backward rapidity ranges ($2.03 < y_{\rm cms} < 3.53$  and $-4.46 <y_{\rm cms}< -2.96$) are 
obtained by studying the \jpsi\ decay to $\mu^+\mu^-$, while the
mid-rapidity region ($-1.37 < y_{\rm cms} < 0.43$) is investigated by measuring the  ${\rm e}^+{\rm e}^-$ decay channel. 
The \pt\ dependence of the \jpsi\ 
production cross section and nuclear modification factor are presented for each of the rapidity intervals, as well as the \jpsi\ mean $p_{\rm T}$ values.
Forward and mid-rapidity results show a suppression of the \jpsi\ yield, with respect to \mbox{pp} collisions, which decreases with increasing \pt. At backward rapidity no significant \jpsi\ suppression is observed. 
Theoretical models including a combination of cold nuclear matter effects such as shadowing and partonic energy loss, are in fair agreement with the data, except at forward rapidity and low transverse momentum. The implications of the \mbox{p--Pb} results for the evaluation of cold nuclear matter effects on \jpsi\ production in \mbox{Pb--Pb} collisions are also discussed.}

\end{abstract}
\end{titlepage}
\setcounter{page}{2}
%
The suppression of charmonia, bound states of $c$ and $\bar{c}$ quarks, and in particular of the J/$\psi$ state, has long been proposed as a signature for the formation of a plasma of quarks and gluons (QGP)~\cite{Matsui:1986dk} in ultrarelativistic nucleus-nucleus collisions. 
However, it was soon realized that charmonium production can also be modified by nuclear effects not necessarily related to QGP formation~\cite{Gerschel:1988wn}.
These so-called cold nuclear matter (CNM) effects can be investigated by studying charmonium production in proton-nucleus (\mbox{p--A}) collisions as confirmed by the analysis of results obtained by  
several fixed-target (SPS \cite{Alessandro:2006jt,Arnaldi:2010ky}, HERA~\cite{Abt:2008ya} and  Tevatron~\cite{Leitch:1999ea}) and collider (RHIC~\cite{Adare:2012qf} and LHC~\cite{Abelev:2013yxa,Aaij:2013zxa}) experiments.
      
Theoretical models have studied the production of charmonium in \mbox{p--A} collisions and the effects of the surrounding cold nuclear medium by introducing various mechanisms which include nuclear shadowing, gluon saturation, energy loss and nuclear absorption. 
Models~\cite{Bodwin:1994jh,Amundson:1996qr,Baier:1981uk} inspired by Quantum ChromoDynamics (QCD) describe charmonium production as a two-step process, with the \ccbar\ pair created in a hard parton scattering, followed by its evolution into a bound state with specific quantum numbers. The pair creation is sensitive to the Parton Distribution Functions (PDFs) in both colliding partners and, at high energy, occurs mainly via gluon fusion.
Although PDFs are known to be modified in a nuclear environment, information on the dependence of such modifications on the fraction $x$ (Bjorken-$x$) of the nucleon momentum carried by the gluons and on the four-momentum squared $Q^2$ transferred in the scattering is still limited~\cite{Eskola:2009uj,deFlorian:2011fp,Hirai:2007sx}. Charmonium production measurements can therefore provide insight into the so-called nuclear shadowing, i.e., on how the nucleon gluon PDFs are modified in a nucleus.\\
Modifications of the initial state of the nucleus are also addressed by approaches assuming that at sufficiently high energies, when the quark pair is produced from a dense gluon system carrying small $x$-values in the nuclear target, a coherent effect known as gluon saturation sets in. Such an effect can be described by the Color Glass Condensate (CGC) effective theory, which is characterized by a saturation momentum scale ($Q^{2}_{s}$). When combined with a specific quarkonium production model~\cite{Kharzeev:2005zr,Fujii:2006ab}, it is able to provide predictions for charmonium production in \mbox{p--A} collisions. 
In the context of shadowing and CGC models, a measurement of the charmonium yield as a function of transverse momentum ($p_{\rm T}$) and rapidity ($y$)  is important as it gives access to specific ranges of values of the gluon $x$ and/or $Q^2$. \\
In addition to these purely initial state effects, both the incoming partons and the \ccbar\ pair propagating through the nucleus  
may lose energy by gluon radiation at the various stages of the charmonium formation process~\cite{Sharma:2012dy}. The interference of gluons radiated before and after the hard production vertex can lead to coherent energy loss effects, expected to induce a modification of the charmonium kinematic distributions~\cite{Arleo:2012rs}.\\
Finally, while travelling through nuclear matter, the evolving \ccbar\ pair or, if crossing times are sufficiently large, the fully formed resonance, may break-up into  open charm meson pairs. Although this mechanism, known as nuclear absorption, plays an important role at lower collision energies~\cite{Arnaldi:2010ky}, at the LHC the contribution of this effect to the production cross section is expected to be small, due to the very short crossing time of the pair through the nuclear environment.  

Understanding the role of the cold nuclear matter effects outlined above is essential to further our knowledge of various aspects of the physics of strong interactions, and it is crucial for the interpretation of the results on charmonium production 
in heavy-ion collisions, where the formation of a QGP is expected. In such a hot and dense deconfined medium the color screening mechanism (the QCD analogue of  the Debye screening in QED) can prevent the formation of the heavy-quark bound states, leading to a suppression of quarkonium production~\cite{Matsui:1986dk}.  
In addition, at LHC energies, the large charm quark density may lead to a
(re)generation of charmonium by (re)combination of charm 
quarks~\cite{BraunMunzinger:2000px,Thews:2000rj} in the QGP phase and/or when the system cools down and the formation of hadrons occurs. This effect enhances charmonium production and is expected to be particularly sizeable at low \pt. 
In heavy-ion collisions, a superposition of hot and cold nuclear matter effects is expected, and a quantitative evaluation of the latter is an important prerequisite for a detailed understanding of the former. At lower energy, both at SPS\cite{Alessandro:2004ap,Arnaldi:2007zz,Arnaldi:2009ph} and RHIC~\cite{Adare:2011yf,Abelev:2009qaa}, a suppression of \jpsi\ production, in addition to the CNM effects estimated from \mbox{p--A}(\mbox{d-A}) collisions, was indeed observed.

A suppression of \jpsi\ production has been measured in \mbox{Pb--Pb} collisions at the LHC~\cite{Abelev:2012rv,Chatrchyan:2012np,Aad:2010aa,Abelev:2013ila,Adam:2015rba}. It was quantified via the nuclear modification factor, i.e., the ratio of the \mbox{Pb--Pb} yields with respect to those measured in \mbox{pp} at the same energy, scaled by the number of binary nucleon-nucleon collisions. The suppression has been found to be stronger at forward rapidity and at high \pt\ \cite{Abelev:2013ila,Adam:2015rba}, in agreement with expectations from (re)combination models. 
Similar to the lower energy experiments, accurate measurements in \mbox{p--A} 
collisions are needed to quantitatively assess the contribution of hot and 
cold nuclear matter effects in \mbox{Pb--Pb}.

The first measurements of inclusive \jpsi\ production in \mbox{p--Pb} collisions at the LHC at \sqrts\ \cite{Abelev:2013yxa,Aaij:2013zxa} 
have shown a sizeable suppression, with respect to binary-scaled \mbox{pp} collisions, at forward rapidity (p-going side) and no suppression at backward rapidity (Pb-going side).
The nuclear modification factors are in fair agreement with models based on nuclear shadowing \cite{Ferreiro:2013pua,Albacete:2013ei}. Calculations including a contribution from coherent energy loss~\cite{Arleo:2012rs} also reproduce the data. Corresponding measurements for the less strongly bound $\psi(2S)$ charmonium state are presented in~\cite{Abelev:2014zpa}.
In addition, an extrapolation to \mbox{Pb--Pb} collisions of the \jpsi\ suppression measured in \mbox{p--Pb} showed
that the effects observed in \mbox{Pb--Pb} cannot be ascribed only to CNM~\cite{Abelev:2013yxa}. 

In this situation, a study of the transverse-momentum dependence of \jpsi\ production at LHC energies for various rapidity regions is particularly interesting in order to: (i) reach a deeper understanding and better quantify  the complicated interplay of CNM effects, which are expected to exhibit a 
well-defined kinematical dependence~\cite{Albacete:2013ei,Arleo:2013zua,Fujii:2013gxa}; (ii) determine if the differential features of the \mbox{Pb--Pb} results that suggest the presence of (re)combination effects are still present when the contribution of CNM is considered.
 
In this paper, we present ALICE results on the transverse-momentum dependence of the inclusive \jpsi\ production in \mbox{p--Pb} collisions at \sqrts, measured in three center-of-mass rapidity ($y_{\rm cms}$) ranges: 
backward ($-4.46 < y_{\rm cms}< -2.96$), mid- ($-1.37 < y_{\rm cms} < 0.43$) and forward ($2.03 < y_{\rm cms} < 3.53$). 
The data are from the 2013 LHC \pPb\ run.

At mid-rapidity, \jpsi\ are reconstructed in the ${\rm e}^{+}{\rm e}^{-}$ decay channel with
the ALICE central barrel detectors, covering the pseudorapidity range
$|\eta_{\rm lab}|<$0.9.
For the backward and forward rapidity  analysis, \jpsi\ are detected, through their $\mu^{+}\mu^{-}$ decay 
channel in the muon spectrometer, in the pseudorapidity range $-4 <\eta_{\rm lab}< -2.5$. 

Due to the energy asymmetry of the LHC beams ($E_{\rm p}= 4$ TeV and $E_{\rm Pb} = 1.58\cdot A_{\rm Pb}$ TeV, where $A_{\rm Pb}$= 208 is the Pb atomic mass number), the nucleon-nucleon center-of-mass is shifted, with respect to the laboratory frame, by $\Delta y = 0.465$  in the direction of the proton beam. Since data were collected in two configurations, interchanging the direction of the proton and the Pb
beams in the LHC, the muon spectrometer acceptance covers the forward and backward $y_{\rm cms}$ regions quoted above, where positive (negative) rapidities refer to the direction of the proton (Pb) beam. In the following, the notation \mbox{p--Pb}  (\mbox{Pb--p}) will refer to the first (second) configuration. 

For the dielectron analysis, the central barrel detectors used for the \jpsi\ reconstruction
are the Inner Tracking System (ITS) \cite{Aamodt:2010aa} and the Time Projection Chamber (TPC) \cite{Alme:2010ke}.
The ITS contains six cylindrical layers of silicon detectors, with the innermost layer at a radius of 3.9~cm with respect to the beam axis and the outermost layer at 43~cm. This detector is used for reconstructing the primary interaction vertex as well as vertices from different interactions and secondary vertices from decays of heavy-flavored particles.
The TPC has a cylindrical geometry with an active volume that extends from 85 to 247~cm in the radial direction and 500~cm longitudinally. It is the main central barrel tracking detector and also provides 
particle identification via the measurement of the specific energy loss 
(${\rm d} E/{\rm d}x$) in the detector gas.

The muon spectrometer~\cite{Aamodt:2008zz} is the main detector used in the dimuon analysis. It consists of a 3 T$\cdot$m dipole magnet, coupled with a tracking and a triggering system. Between the interaction point and the muon spectrometer, a ten interaction-length ($\lambda_{\rm I}$) front absorber filters out the hadrons produced in the interaction. Muon tracking is performed by means of five tracking stations, each one made of two planes of Cathode Pad Chambers.
A 7.2 $\lambda_{\rm I}$ iron wall, which stops secondary hadrons escaping the front absorber and low momentum muons, is placed after the tracking stations. It is followed by a muon trigger system, based on two stations equipped with  Resistive Plate Chambers.
A conical absorber made of tungsten, lead and steel protects the spectrometer against secondary particles produced by the interaction of \mbox{large-$\eta$} primary particles in the beam pipe.
In the dimuon analysis, the determination of the interaction vertex is provided by the two innermost Si-pixel layers of the ITS (Silicon Pixel Detector, SPD). 

For both analyses, timing information from the Zero Degree Calorimeters~\cite{ALICE:2012aa}, placed symmetrically at 112.5 m with respect to the interaction point, is used to remove de-bunched proton-lead collisions.
Furthermore, two scintillator hodoscopes (VZERO)~\cite{Abbas:2013taa}, with pseudorapidity coverage $2.8 <\eta_{\rm lab}< 5.1$ and 
$-3.7 <\eta_{\rm lab} < -1.7$, are used to remove beam-induced background.
More details on the ALICE apparatus can be found in \cite{Aamodt:2008zz}.

A coincidence of signals in the two VZERO detectors provides the minimum bias (MB) trigger, which has a $>99$\% efficiency for 
selecting non single-diffractive \mbox{p--Pb} collisions~\cite{ALICE:2012xs}. While the dielectron analysis is based on MB-triggered events, the study of \jpsi\ in the $\mu^{+}\mu^{-}$ decay channel relies on a dimuon trigger which 
requires, in addition to the MB condition, the detection of two opposite-sign tracks in the trigger system. 
The dimuon trigger selects two muon candidates with transverse momenta $p_{{\mathrm T},\mu}$ larger than 0.5 GeV/$c$. The trigger threshold is not sharp, and the single muon trigger efficiency reaches its plateau value ($\sim$ 96\%) at $p_{{\mathrm T},\mu}\sim1.5$ GeV/$c$. 
The dielectron analysis was performed on a data sample corresponding to the $\mbox{p--Pb}$ configuration, with an integrated luminosity $\mathcal{L}_{\rm int}= 51.4 \pm 1.9$ $\mu$b$^{-1}$, while for the 
dimuon analysis the corresponding values are $5.01 \pm 0.19$ nb$^{-1}$ for 
\mbox{p--Pb} and $5.81 \pm 0.20$ nb$^{-1}$ for \mbox{Pb--p} (the quoted uncertainties are systematic)~\cite{Abelev:2014epa}. 

\begin{figure}[hbtp]
\centering
\resizebox{0.8\textwidth}{!}
{\includegraphics{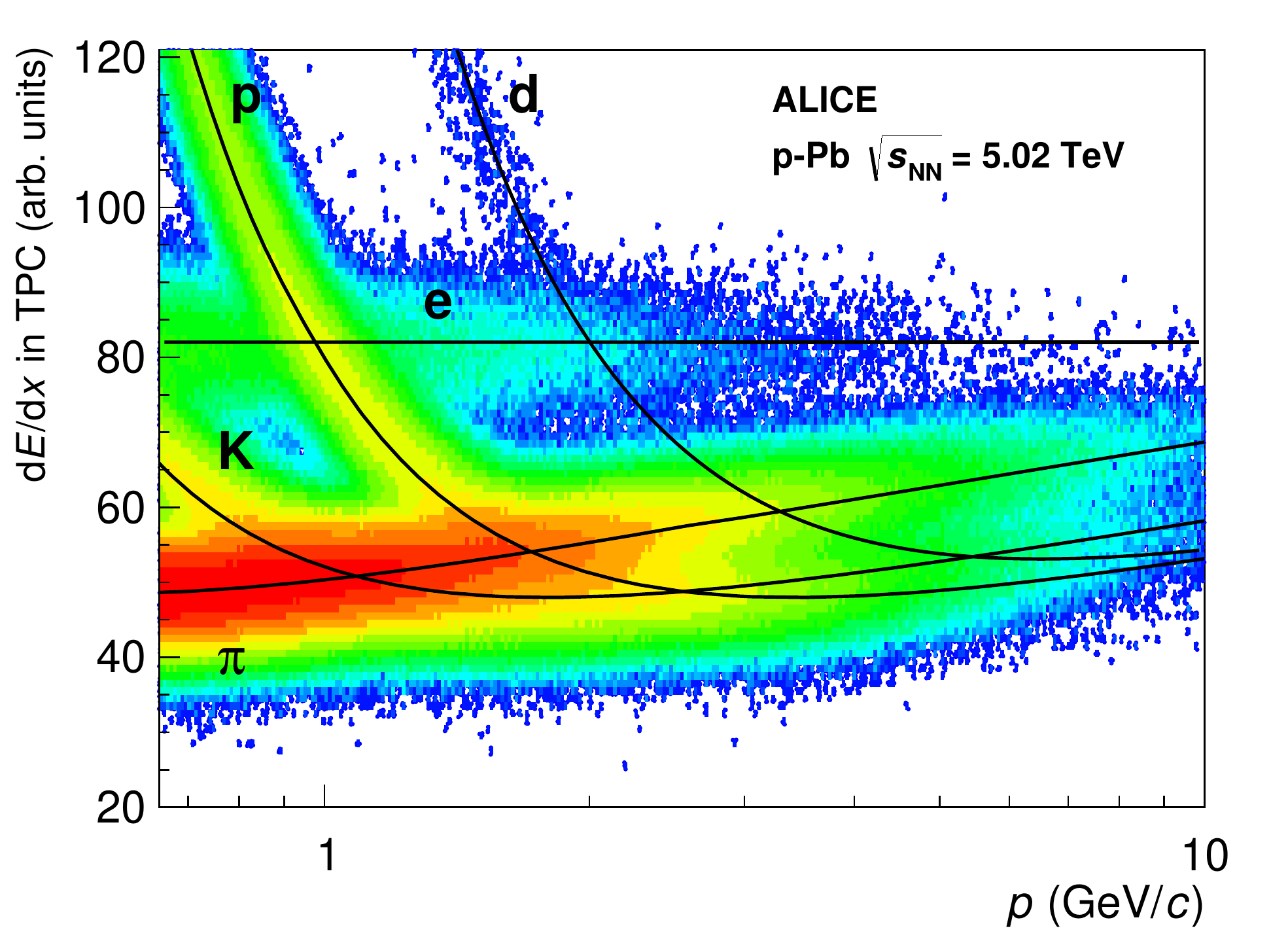}}
\caption{Charged particle specific energy loss (${\rm d} E/{\rm d}x$) as a function of momentum, as measured in the TPC in \mbox{p--Pb} collisions. 
The black lines are the corresponding Bethe-Bloch parametrizations for the various particle species.}
\label{fig:dEdx}
\end{figure}

The dielectron analysis is based on 1.07$\times10^8$ events, collected with a low MB interaction rate ($\sim$10 kHz), with a negligible amount of events having more than one interaction per bunch crossing (pile-up events). The interaction vertex is required to lie within $\pm 10$ cm from the nominal collision point along the beam axis, in order to obtain a uniform acceptance of the central barrel detector system in the fiducial range 
$|\eta_{\rm lab}|< 0.9$.
Electron candidates are selected with criteria 
very similar to those used in previous analyses of pp collisions at $\sqrt{s}=$7~TeV \cite{Aamodt:2011gj} and Pb--Pb collisions
at $\sqrt{s_{\rm NN}} = 2.76$~TeV \cite{Abelev:2013ila}. To ensure a uniform tracking efficiency and particle identification resolution in the TPC, only tracks 
within $|\eta_{\rm lab}|<0.9$ are used. Electron identification is performed using the TPC, as shown in Fig.~\ref{fig:dEdx}, by requiring the ${\rm d}E/{\rm d}x$ signal to be compatible with the electron assumption within 3$\sigma$, where $\sigma$ denotes the resolution of the ${\rm d}E/{\rm d}x$ measurement. Furthermore, the TPC tracks that are compatible with the pion and proton assumptions within 3.5$\sigma$ 
are rejected. A slightly looser rejection condition (3$\sigma$) is applied when considering tracks corresponding to dielectron candidates with $p_{\rm T}>$ 5 GeV/$c$ in order to enhance the statistics. 
A cut on the transverse momentum ($p_{\rm T,e}>1.0$~GeV/$c$) is applied to remove combinatorial background from low-momentum electrons. The efficiency loss induced by this cut amounts to only $\sim$20\%, due to the relatively large 
momentum of the \jpsi\ decay products. The electron candidates must have at least one hit in the innermost two layers of the ITS, thus rejecting a large fraction of background electrons from photon conversions. For dielectrons with $p_{\rm T}<3$ GeV/$c$ the electron candidates are required to have a hit in the first layer, to further reduce background. 
The tracks are required to have at least 70 out of a maximum of 159 clusters in the TPC and a $\chi^2$ normalized to the number of clusters attached to the track 
smaller than 4.  

\begin{figure}[h!]
\begin{center}
\includegraphics[width=1.0\textwidth]{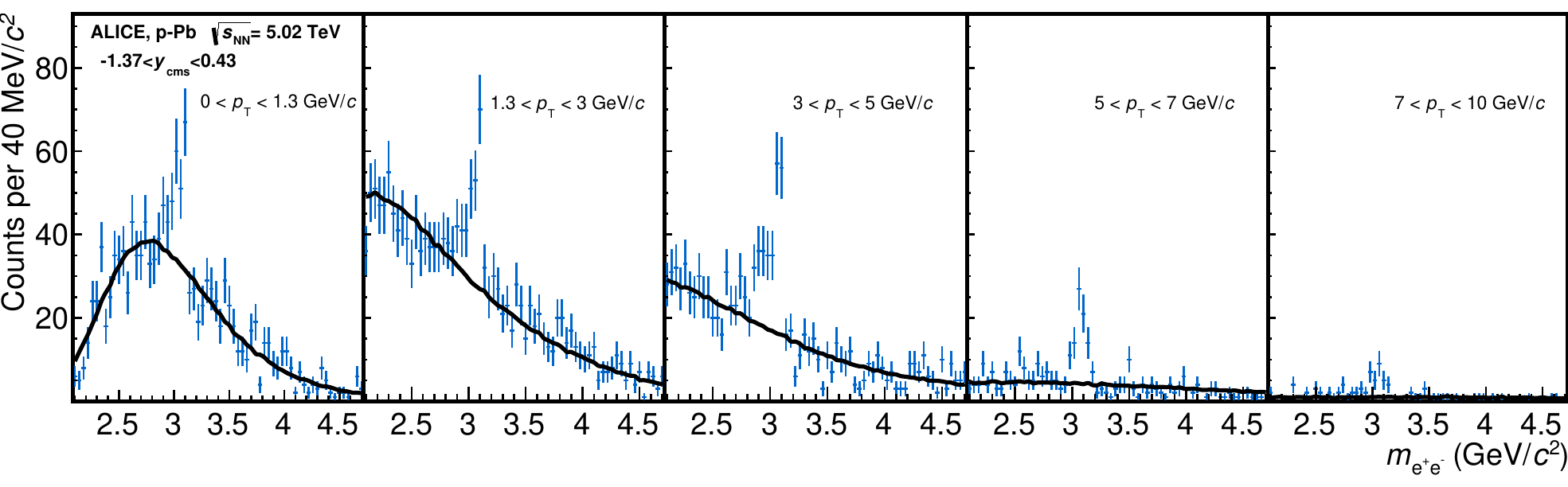}
\end{center}
\caption{Opposite-sign dielectron invariant mass 
spectra (blue symbols) for various $p_{\rm T}$ intervals, compared to the background (black curve) estimated through mixed events. 
The background is scaled to match the data in the mass ranges $2.0<m_{{\rm e}^+{\rm e}^-}<2.5$~GeV/$c^2$ and  $3.2<m_{{\rm e}^+{\rm e}^-}<3.7$~GeV/$c^2$.}
\label{fig:invmassee}
\end{figure}

The \jpsi\ yields are obtained by counting the number of entries in the invariant mass range $2.92<m_{{\rm e}^+{\rm e}^-}<3.16$~GeV/$c^2$ after background subtraction. The \jpsi\ radiative decay channel and the
energy loss of the electrons due to bremsstrahlung in the detector material produce a long tail towards low invariant
masses. A fit using a Crystal Ball (CB)~\cite{Gaiser} function for the \jpsi\ signal gives compatible values in Monte-Carlo (MC) and data ($\sim$20 MeV/$c^2$ for the width of the Gaussian component of the CB).
Taking into account such a mass resolution and the presence of the bremsstrahlung tail, $67-73\%$ of the signal, depending on \pt, falls within the counting window. 
The background shape is obtained from event mixing. Event mixing is performed by pairing leptons from different events having similar global characteristics such as the primary-vertex position and the 
track multiplicity (the result being quite insensitive to the rapidity range, either forward or central, chosen for the multiplicity measurements).
The mixed-event background is then scaled to match the 
same-event opposite-sign distribution in the mass ranges $2.0<m_{{\rm e}^+{\rm e}^-}<2.5$~GeV/$c^2$ and $3.2<m_{{\rm e}^+{\rm e}^-}<3.7$~GeV/$c^2$ (the contribution of the bremsstrahlung tail in the former range and of the $\psi(2S)$ in the latter are negligible).  Consistent results are found when the same-event like-sign distributions are used, instead 
of event mixing, to estimate the 
background. The systematic uncertainty on the signal extraction comes from the variation of the mass range where the normalization of the mixed-event background shape is performed and from the choice of the mass window where the signal is counted. 
The signal extraction has been performed in five transverse-momentum bins, $p_{\rm T} < 1.3$, $1.3 < p_{\rm T}< 3$, $3 < p_{\rm T} < 5$, $5 < p_{\rm T} < 7$ and $7 < p_{\rm T} <10$ GeV/$c$.  The \jpsi\ counts in these bins vary from 25 to 132, with a significance, computed in the $2.92<m_{{\rm e}^+{\rm e}^-}<3.16$~GeV/$c^2$ mass region, ranging from 4.6 to 8.7. An analysis of the \pt-integrated data sample, using the procedure detailed above, gives $465\pm37 ({\rm stat.}) \pm 16 ({\rm syst.})$ \jpsi\ signal counts. The systematic uncertainty on the signal extraction is largest at low \pt\  (10\% for $p_{\rm T} < 1.3$ GeV/$c$ and 12\% for $1.3 < p_{\rm T}< 3$ GeV/$c$), due to a less favorable signal over background ratio, and decreases to $\sim$5.5--8.4\% in the other three \pt\ bins. 
Figure \ref{fig:invmassee} shows the invariant mass distributions for the opposite-sign dielectrons compared with the mixed-event background for the different intervals of \pt. 

\begin{figure}[htbp]
\centering
\resizebox{1.0\textwidth}{!}
{\includegraphics{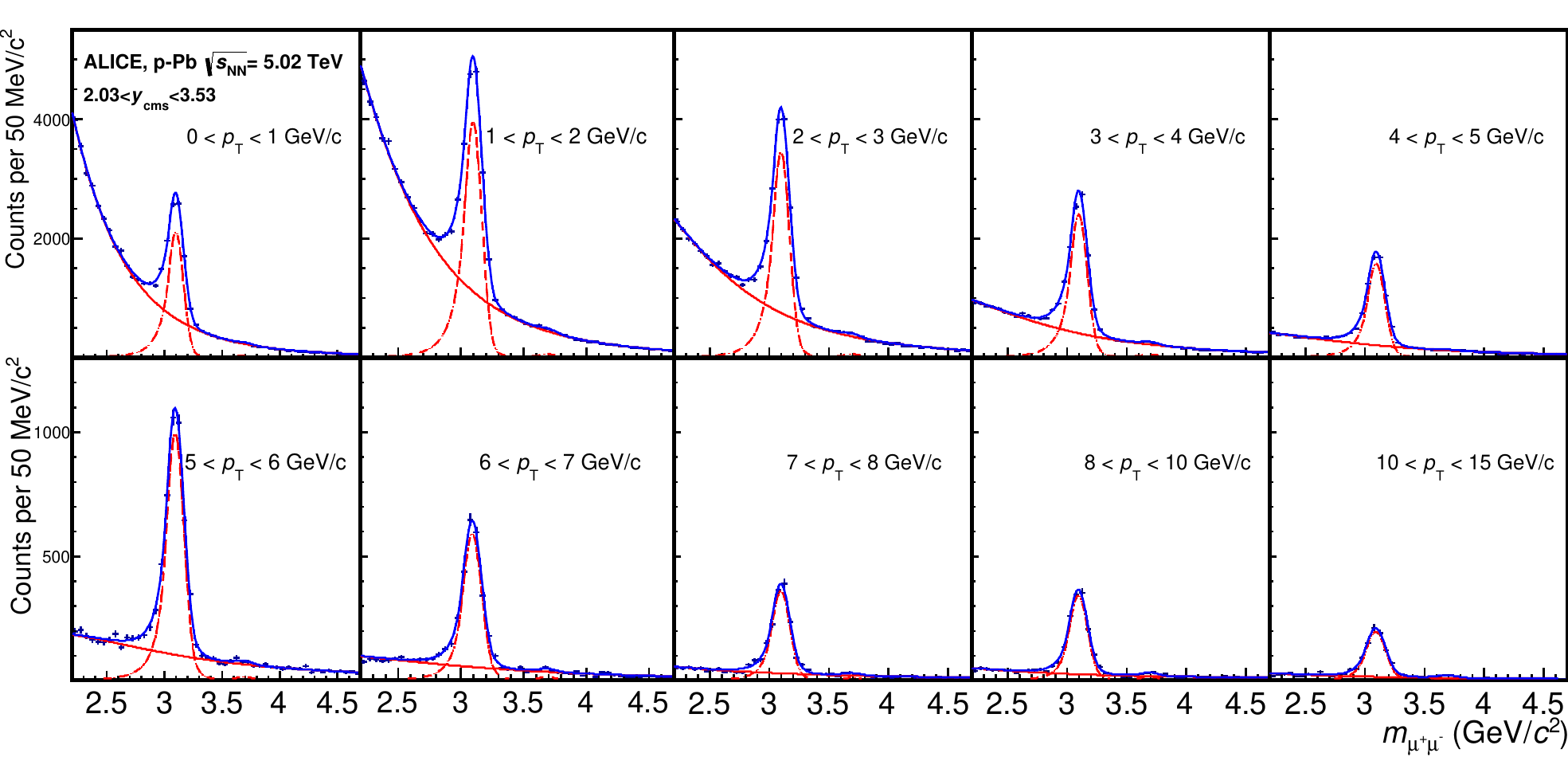}}
\caption{ The opposite-sign dimuon invariant mass spectra for the  various \pt\ bins, relative to the \mbox{p--Pb} data sample (blue symbols). 
The fits shown in this Figure (blue curves) were performed by using the sum of extended Crystal Ball functions for the \jpsi\ and 
\psip\ signals, and a variable width Gaussian for the background. The signal and background components are shown separately as red curves.}
\label{fig:invmasspA}
\end{figure}

The dimuon analysis is performed as detailed in \cite{Abelev:2013yxa}, and is shortly summarized hereafter.
Data were collected with the dimuon trigger, and the MB interaction rate (up to 200 kHz) was much higher than in the sample used for the dielectron analysis. This leads to a $\sim$2\% interaction pile-up probability. However, the probability of having more than one dimuon in the same bunch crossing satisfying the trigger condition is negligible. 
Muon candidate tracks are reconstructed in the tracking system by using the standard reconstruction algorithm~\cite{Aamodt:2011gj}.
The quality of the tracks is ensured by requiring the single muon pseudorapidity to be in the range $-4 <\eta_{{\rm lab,}\mu} < -2.5$, in order to remove particles at the edges of the muon spectrometer acceptance. In addition, a cut on the radial coordinate of the track at the end of the front absorber ($17.6<R_{\rm abs}<89.5$ cm) is performed, ensuring rejection of muons crossing its high-density part, where energy loss and multiple scattering effects are more important. The tracks reconstructed in the tracking system that are not matched to a corresponding track in the triggering system are rejected~\cite{Aamodt:2011gj}. Finally, the reconstructed dimuons are required to be in  $2.03 < y_{\rm cms} < 3.53$ ($-4.46 <y_{\rm cms}< -2.96$) for the forward (backward) rapidity analysis. 
The number of \jpsi\ is extracted in transverse-momentum bins, in the range $p_{\rm T} < 15$ GeV/$c$, through fits to the invariant mass 
spectra of opposite-sign dimuons. The spectra are fitted with a superposition of background and resonance shapes. The background is described with a Gaussian function with a mass-dependent width or, alternatively, with an exponential function times a fourth-order polynomial function. 
For the \jpsi\ shape an extended Crystal Ball function, which accommodates a non-Gaussian tail both on the right and on the left side of the resonance peak, is adopted. Alternatively, a pseudo-Gaussian function~\cite{Ruben} is used, corresponding to a Gaussian core around the 
\jpsi\ pole, and tails on the right and left side of it, parameterized by varying the width of the Gaussian as a function of the mass.
The value of the \jpsi\ mass and its width ($\sigma$) at the pole position are free parameters of the fit. The
mass coincides with the PDG value within less than 5 MeV/$c^2$ and the width is  $\sim$70 MeV/$c^{2}$, slightly increasing with \pt, due to a small relative decrease in the tracking resolution for harder muons.
Although the signal over background ratios, calculated for a $\pm 3\sigma$ interval around the resonance peak, are relatively large (ranging from 1.4 to $\sim 6$ moving from low to high \pt), the parameters of the tails of the \jpsi\ distributions cannot be reliably tuned on the data (in particular at large \pt, where statistics is limited), but are fixed, for each \pt\ bin, to the values extracted from fits to reconstructed samples from a signal-only MC generation. The contribution of the \psip\ resonance is also included in the fitting procedure, even if its influence on the determination of the \jpsi\ yield is negligible. Finally, all the fits are performed in two different invariant mass ranges, either $2<m_{\mu\mu}<5$ GeV/$c^{2}$ or $2.2<m_{\mu\mu}<4.5$ GeV/$c^{2}$.
Examples of fits to the invariant mass spectra, in the \pt\ bins under study, are shown in Fig.~\ref{fig:invmasspA}.
 
For each \pt\ bin, the number of \jpsi\ is evaluated as the average of the integrals of the resonance functions obtained in the various fits. The RMS of the corresponding yield distributions ($0.2-3$\%, depending on \pt) provides the systematic uncertainty on the signal extraction. Additional sets of tails, obtained from the MC, but referring to other $y_{\rm cms}$ and $p_{\rm T}$ phase space regions, have also been tested and the dependence of the extracted yields on the variation of the tails (2\%) is included in the systematic uncertainty on the signal extraction.
As a function of \pt, the number of \jpsi\ in the \mbox{p--Pb} (\mbox{Pb--p}) configuration ranges between $\sim$16100 ($\sim$16000) in the most populated bin ($1 < p_{\rm T} < 2$  GeV/$c$) and less than $\sim$900 ($\sim$300) in the highest \pt\ bin ($10 < p_{\rm T} < 15$  GeV/$c$). 

The \jpsi\ yields are then corrected for the product of acceptance times  efficiency ($A\times\varepsilon$), evaluated by means of a MC simulation. 
\jpsi\ production is assumed to be unpolarized, as motivated by the small degree of polarization measured in \mbox{pp} collisions at $\sqrt{s}= 7$ TeV~\cite{Abelev:2011md,Aaij:2011jh,Chatrchyan:2013cla}. 
In the ${\rm e}^+{\rm e}^-$ decay channel, $A\times\varepsilon$ is calculated using a MC simulation where \jpsi\ are injected into \mbox{p--Pb} collisions simulated with HIJING~\cite{Wang:1991hta}. The decay products of the \jpsi\ are then propagated through a realistic 
description of the ALICE set-up, based on GEANT3.21~\cite{Brun5013}, taking into account the time evolution of the detector performance. Finally, \jpsi\ candidates are reconstructed with the same procedure applied to data.
The \pt-integrated $A\times\varepsilon$ factor amounts to 8.9\%. Its \pt-dependence exhibits a minimum ($\sim$7.5\%) around $p_{\rm T}=2$~GeV/$c$, due to the kinematical acceptance, and it reaches $\sim$12\% at high \pt. The integrated value of $A\times\varepsilon$ is affected by a 3\% systematic uncertainty related to the choice of the \jpsi\ \pt- and $y$-distributions used in the MC simulation. This value is obtained using as input  several distributions, determined by varying within uncertainties the differential spectra extracted from the ALICE \mbox{p--Pb} data themselves. For \pt-differential studies, the values of $A\times\varepsilon$ are found to be sensitive only at a sub-percent level to the adopted input \pt- and $y$-distributions. 
A further small systematic uncertainty reaching 1.5\%  in the highest \pt\ interval and related to the statistical uncertainty of the MC sample is also introduced.
The systematic uncertainty on the dielectron reconstruction efficiency is strongly dominated by the particle identification uncertainty and amounts to 4\%. It was obtained by comparing the single track reconstruction efficiency for topologically identified positrons and electrons from photon conversions with the corresponding MC quantities. 
In the dimuon analysis, the \jpsi\ $A\times\varepsilon$ is obtained with a MC simulation, by generating signal-only samples, tracking them in the experimental set-up modeled with GEANT3.21 and using the same reconstruction procedure applied to data. The use of a pure signal MC is justified, since  the tracking efficiency does not show a dependence on the hadronic multiplicity of the collision.  
A realistic description of the set-up is adopted, including the time evolution of the efficiencies of tracking and triggering detectors. As for the dielectron analysis, the differential distributions used as an input to the MC are tuned directly on the data. 
The \jpsi\ $A\times\varepsilon$ values, integrated over \pt, are 25.4\% and 17.1\% for \mbox{p--Pb} and \mbox{Pb--p} respectively~\cite{Abelev:2013yxa}, and exhibit a dependence on transverse momentum, being of the order of $\sim$24\%  ($\sim$16\%) for \mbox{p--Pb} (\mbox{Pb--p}) at low \pt\ and reaches $\sim$50\% ($\sim$35\%) in the highest \pt\ bin ($10 < p_{\rm T} < 15$ GeV/$c$). The systematically lower $A\times\varepsilon$ values in \mbox{Pb--p} reflect the smaller detector efficiency in the corresponding data taking period. The systematic uncertainty on the integrated $A\times\varepsilon$ due to the input shapes is 1.5\% for both \mbox{p--Pb} and \mbox{Pb--p}, and has been estimated using various distributions obtained from data and corresponding to smaller intervals in $y$, \pt\ and centrality (see~\cite{Abelev:2013yxa} for details). For \pt-differential studies, the corresponding uncertainties are below 1.5\%.
The uncertainty on the dimuon tracking efficiency amounts to 4\% (6\%) for
\mbox{p--Pb} (\mbox{Pb--p}) and is taken as constant for the full \pt\ range. It is evaluated by combining the uncertainties on single muon tracking efficiencies, considered as uncorrelated. 
The efficiency of each tracking plane is obtained using the redundancy of the tracking system (two independent planes per station) and then single muon efficiencies for the full tracking system 
are calculated according to the tracking algorithm~\cite{Aphecetche09}. 
Their uncertainty is determined by comparing the efficiency obtained with tracks from MC and real data.
The systematic uncertainty on the dimuon trigger efficiency includes: (i) a 
contribution due to the uncertainty in the evaluation of the trigger detector 
efficiency ($\sim$ 2\%, independent of \pt); (ii) a $0.5-3$\% \pt-dependent contribution (2\% for the integrated efficiency), related to small differences in the trigger response function between data and MC in the region close to the trigger threshold; (iii) a $0.5-3.5$\% \pt-dependent contribution due to a 
small fraction of opposite-sign pairs which were misidentified as like-sign by the trigger system. Finally, a $\sim$1\% uncertainty, independent of \pt, is included, due to the choice of the value of the $\chi^{2}$ cut applied to the matching of tracks reconstructed in the muon tracking and triggering systems.


\begin{table}
\centering
\begin{tabular}{l|c|c|c}
\hline
Source & $\sigma^{{\rm J/}\psi}_{\rm pPb}$, $R_{\rm pPb}$ & $\sigma^{{\rm J/}\psi}_{\rm pPb}$, $R_{\rm pPb}$ & $\sigma^{{\rm J/}\psi}_{\rm Pbp}$, $R_{\rm Pbp}$ \\ 
 & -1.37$<y_{\rm cms}<$0.43& 2.03$<y_{\rm cms}<$3.53 &  -4.46$<y_{\rm cms}<$-2.96  \\ \hline
{\it Uncorrelated} & & & \\
Tracking efficiency ($\mu^+\mu^-$) & - & 4 & 6\\
Trigger efficiency ($\mu^+\mu^-$) & - & 2.7 $-$ 4.1 & 2.7 $-$ 4.1\\
Matching efficiency ($\mu^+\mu^-$)& - & 1 & 1 \\ 
Reconstruction efficiency (${\rm e}^+{\rm e}^-)$ & 4 & - & -\\
Signal extraction & 5.5 $-$ 12.6 & 2 $-$ 2.5 & 2 $-$ 3.6\\
MC input & 0.3 $-$ 1.5 & 0.1 $-$ 0.4 & 0.1 $-$ 1.4\\
$\sigma^{{\rm J/}\psi}_{\rm pp}$ & 4.8 $-$ 15.7 & 5.2 $-$ 9.2 & 5.2 $-$ 9.2\\ \hline
{\it Partially correlated} & & \\
$\sigma^{{\rm J/}\psi}_{\rm pp}$ (corr. vs $y$ and \pt) & - & 2.8 $-$ 5.9 & 2 $-$ 5.6 \\ \hline
{\it Correlated} & &  \\
B.R. (\jpsi\ $\rightarrow l^{+}l^{-}$) & 1 & 1& 1  \\
$\mathcal{L}_{int}$(corr. vs. \pt, uncorr. vs. $y$) & 3.3 & 3.4 & 3.1\\
$\mathcal{L}_{int}$(corr. vs. $y$ and \pt) & 1.6 & 1.6 & 1.6\\
$\sigma^{{\rm J/}\psi}_{\rm pp}$ & 16.6 & 5.2 & 5.2 \\ \hline
\end{tabular}
\caption{Systematic uncertainties (in percent) on the measurement of inclusive J/$\psi$ cross sections and nuclear modification
factors. For \pt-dependent uncertainties, the minimum and
maximum values are given. The degree of correlation (uncorrelated, partially correlated, correlated) refers to the $p_{\rm T}$-dependence, unless specified otherwise.
It cannot be excluded that a  
degree of correlation, difficult to quantify, is present also in uncertainties currently labelled as uncorrelated. 
Uncertainties on $\mathcal{L}_{int}$ and branching ratios are relevant for cross sections, while those on $\sigma^{{\rm J/}\psi}_{\rm pp}$ 
contribute only to the uncertainty on the nuclear modification factors. $\mathcal{L}_{int}$ uncertainties are split into two components, 
respectively uncorrelated and correlated between \mbox{p--Pb} and \mbox{Pb--p}, as detailed in~\cite{Abelev:2014epa}.}
\label{tab:syst}
\end{table}

\begin{figure}[h!]
\centering
\resizebox{0.8\textwidth}{!}
{\includegraphics{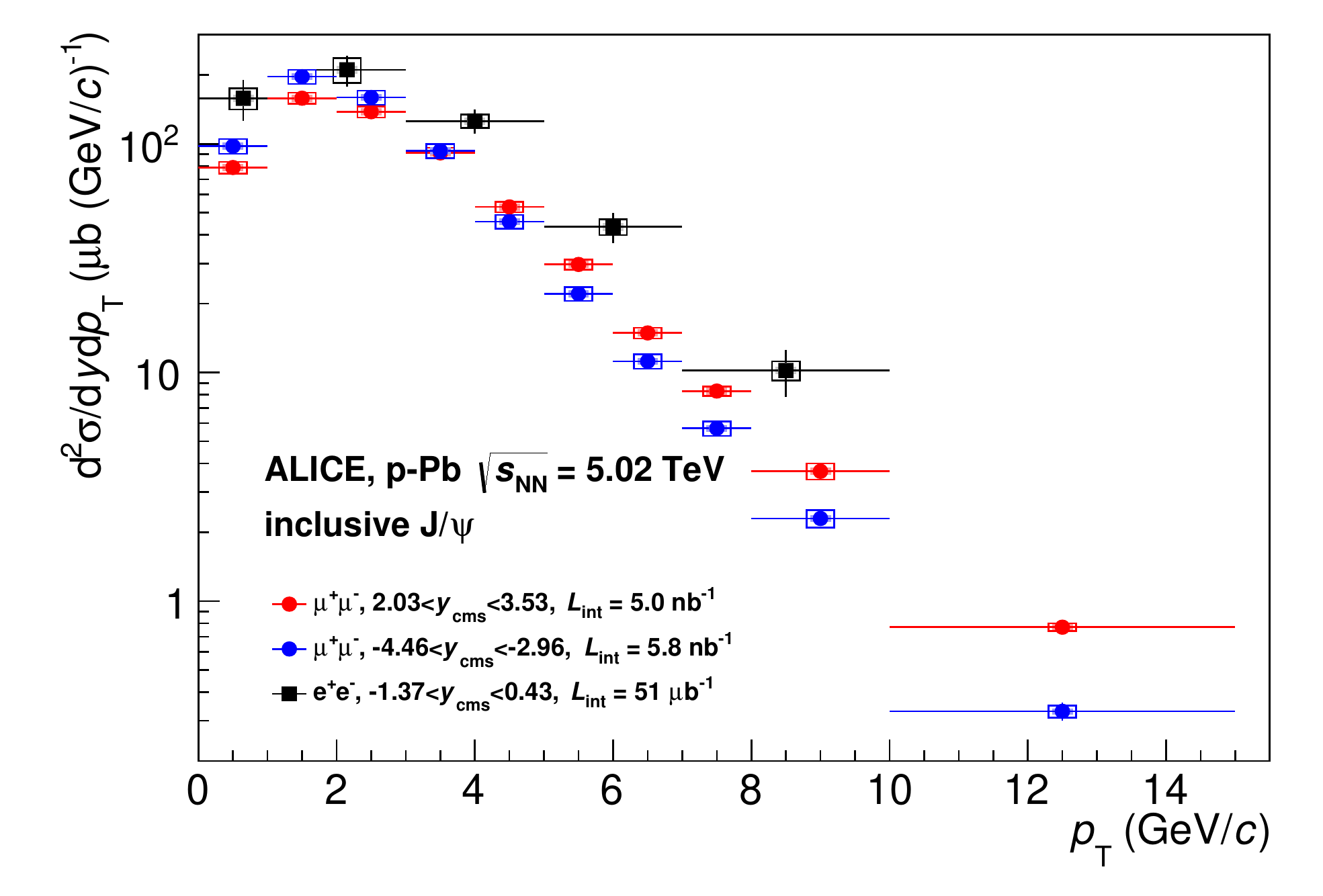}}
\caption{\pt-differential inclusive \jpsi\ cross sections for the various rapidity regions under study. 
The vertical error bars correspond
to the statistical uncertainties, while open boxes represent the uncorrelated
systematic uncertainties and the shaded boxes the quadratic sum of the fully and partially correlated ones. The numerical values can be read in Table~\ref{tab:2}.
The horizontal bars correspond to the widths of the $p_{\rm T}$ bins.}
\label{fig:crosssections_vs_pt}
\end{figure}

The differential cross section for inclusive J/$\psi$ production is defined as:

\begin{equation}
\frac{{\rm d}^{2}\sigma^{\rm J/\psi}_{\rm pPb}}{{\rm d}y{\rm d}p_{\rm T}} =\frac{N_{\rm J/\psi}(\Delta y,\Delta p_{\rm T})}
{\mathcal{L}_{\rm int}^{\rm pPb} \cdot (A\times\varepsilon)_{(\Delta y,\Delta p_{\rm T})} \cdot {\rm B.R.}({\rm J}/\psi\rightarrow l{^+}l{^-}) \cdot \Delta y \cdot \Delta p_{\rm T}}
\end{equation}

where $N_{\rm J/\psi}(\Delta y,\Delta p_{\rm T})$ is the number of J/$\psi$ for a given $\Delta y$ and $\Delta$\pt\ interval.
The branching ratio to dileptons, ${\rm B.R.}({\rm J}/\psi\rightarrow l{^+}l{^-})$, is 
$5.94\pm 0.06$\% ($5.93\pm 0.06$\%) for the dielectron (dimuon) decay~\cite{Beringer:1900zz}. 
The integrated luminosity, $\mathcal{L}_{\rm int}^{\rm pPb}$, is the ratio between $N_{\rm MB}$, the number of MB collisions, 
and $\sigma^{\rm MB}_{\rm pPb}$, the corresponding cross section, measured in a van der Meer scan to be 2.09 $\pm$ 0.07 b for the \mbox{p--Pb} configuration and 
2.12 $\pm$ 0.07 b for the \mbox{Pb--p} case~\cite{Abelev:2014epa}. The luminosity is also independently determined by means of a second signal 
based on a \v{C}herenkov counter~\cite{Aamodt:2008zz}, as
described in~\cite{Abelev:2014epa}. The two measurements differ by at most 1\% throughout the whole data-taking period
and such a value is quadratically added to the luminosity uncertainty.
Finally, since the dimuon analysis is based on a sample of $N_{\rm DIMU}$ dimuon triggered events, the number of 
equivalent MB collisions is computed as $N_{\rm MB}=F \cdot N_{\rm DIMU}$, 
where $F$ is a factor accounting for the probability of having a dimuon
trigger when the MB condition is satisfied and for the small ($\sim2\%$) 
pile-up probability in the corresponding data sample. The systematic uncertainty on this quantity, quadratically added to the other luminosity uncertainties, 
is 1\% and originates from the comparison between the different approaches used for its evaluation~\cite{Abelev:2013yxa}.
A summary of the systematic uncertainties can be found in Table~\ref{tab:syst}.
The differential inclusive \jpsi\ cross sections are shown in
Fig.~\ref{fig:crosssections_vs_pt}, in the ranges $p_{\rm T}<10$ GeV/$c$ for the dielectron analysis and $p_{\rm T}<15$ GeV/$c$ for the dimuon analysis. The numerical values can be found in Table~\ref{tab:2}.

\begin{table}
\centering
\small
\begin{tabular}{cc|ccc}
$p_{\rm T}$ & $\mathrm{d}^2\sigma^{\rm J/\psi}_{\rm pPb}/\mathrm{d} y\mathrm{d}p_{\rm T}$ & $p_{\rm T}$ & $R_{\rm pPb}$ 
& $\mathrm{d}^2\sigma^{\rm J/\psi}_{\rm pp}/\mathrm{d} y\mathrm{d}p_{\rm T}$ (interpol.)\\
 (GeV/$c$) & ($\mu$b/(GeV/$c$)) & (GeV/$c$) & & ($\mu$b/(GeV/$c$)) \\
\hline
 \multicolumn{5}{c}{$-4.46<y_{\rm cms}<-2.96$ ($\mu^{+}\mu^{-}$)} \\  
\hline
 $[0; 1]$ & 97.7$\pm$2.0$\pm$7.2$\pm$3.5   & $[0; 1]$ & 0.96$\pm$0.02$\pm$0.09$\pm$0.03$\pm$0.06 & 0.490$\pm$0.029$\pm$0.017$\pm$0.026\\
 $[1; 2]$ & 196.8$\pm$2.7$\pm$14.3$\pm$7.1 & $[1; 2]$ & 1.06$\pm$0.01$\pm$0.10$\pm$0.04$\pm$0.07 & 0.892$\pm$0.048$\pm$0.030$\pm$0.046\\
 $[2; 3]$ & 159.6$\pm$2.1$\pm$11.6$\pm$5.8 & $[2; 3]$ & 1.11$\pm$0.01$\pm$0.10$\pm$0.04$\pm$0.07 & 0.693$\pm$0.036$\pm$0.025$\pm$0.036\\
 $[3; 4]$ & 93.3$\pm$1.6$\pm$6.7$\pm$3.4   & $[3; 4]$ & 1.16$\pm$0.02$\pm$0.10$\pm$0.04$\pm$0.07 & 0.388$\pm$0.021$\pm$0.012$\pm$0.020\\
 $[4; 5]$ & 45.7$\pm$1.0$\pm$3.2$\pm$1.7   & $[4; 5]$ & 1.17$\pm$0.02$\pm$0.11$\pm$0.03$\pm$0.07 & 0.187$\pm$0.011$\pm$0.004$\pm$0.010\\
 $[5; 6]$ & 22.1$\pm$0.5$\pm$1.6$\pm$0.8   & $[5; 6]$ & 1.13$\pm$0.03$\pm$0.12$\pm$0.02$\pm$0.07 & 0.094$\pm$0.007$\pm$0.002$\pm$0.005\\
 $[6; 7]$ & 11.2$\pm$0.4$\pm$0.8$\pm$0.4   & $[6; 8]$ & 1.27$\pm$0.03$\pm$0.14$\pm$0.08$\pm$0.08 & 0.032$\pm$0.003$\pm$0.002$\pm$0.002\\
 $[7; 8]$ & 5.7$\pm$0.3$\pm$0.4$\pm$0.2 &  &  &  \\
 $[8; 10]$ & 2.3$\pm$0.1$\pm$0.2$\pm$0.1 &  &  &  \\
 $[10; 15]$ & 0.33$\pm$0.03$\pm$0.03$\pm$0.01 &  &  &  \\
\hline
 \multicolumn{5}{c}{$-1.37<y_{\rm cms}<0.43$ (${\rm e}^{+}{\rm e}^{-}$)} \\  
\hline
 $[0; 1.3]$ & 158$\pm$33$\pm$17$\pm$6 & $[0; 1.3]$ & 0.81$\pm$0.17$\pm$0.10$\pm$0.14 & 0.94$\pm$0.07$\pm$0.16\\
 $[1.3; 3]$ & 211$\pm$33$\pm$26$\pm$8 & $[1.3; 3]$ & 0.64$\pm$0.10$\pm$0.09$\pm$0.11 & 1.60$\pm$0.08$\pm$0.26\\
 $[3; 5]$ & 126$\pm$15$\pm$9$\pm$5 & $[3; 5]$ & 0.77$\pm$0.09$\pm$0.07$\pm$0.13 & 0.79$\pm$0.05$\pm$0.13\\
 $[5; 7]$ & 43.4$\pm$6.5$\pm$3.4$\pm$1.7 & $[5; 7]$ & 0.89$\pm$0.13$\pm$0.13$\pm$0.15 & 0.23$\pm$0.03$\pm$0.04\\
 $[7; 10]$ & 10.2$\pm$2.4$\pm$1.0$\pm$0.4 & $[7; 10]$ & 0.89$\pm$0.21$\pm$0.16$\pm$0.15 & 0.06$\pm$0.01$\pm$0.01\\
\hline
 \multicolumn{5}{c}{$2.03<y_{\rm cms}<3.53$ ($\mu^{+}\mu^{-}$)}\\  
\hline
 $[0; 1]$ & 78.8$\pm$1.5$\pm$4.6$\pm$3.1 & $[0; 1]$ & 0.61$\pm$0.01$\pm$0.05$\pm$0.02$\pm$0.04 & 0.624$\pm$0.036$\pm$0.025$\pm$0.032\\
 $[1; 2]$ & 158.4$\pm$2.2$\pm$9.0$\pm$6.2 & $[1; 2]$ & 0.64$\pm$0.01$\pm$0.05$\pm$0.02$\pm$0.04 & 1.197$\pm$0.064$\pm$0.046$\pm$0.062\\
 $[2; 3]$ & 138.2$\pm$1.9$\pm$7.9$\pm$5.4 & $[2; 3]$ & 0.68$\pm$0.01$\pm$0.05$\pm$0.03$\pm$0.04 & 0.980$\pm$0.051$\pm$0.039$\pm$0.051\\
 $[3; 4]$ & 91.3$\pm$1.4$\pm$5.0$\pm$3.6  & $[3; 4]$ & 0.76$\pm$0.01$\pm$0.06$\pm$0.03$\pm$0.05 & 0.579$\pm$0.032$\pm$0.022$\pm$0.030\\
 $[4; 5]$ & 53.0$\pm$0.9$\pm$2.8$\pm$2.1  & $[4; 5]$ & 0.87$\pm$0.02$\pm$0.07$\pm$0.02$\pm$0.06 & 0.294$\pm$0.017$\pm$0.008$\pm$0.015\\
 $[5; 6]$ & 29.7$\pm$0.6$\pm$1.6$\pm$1.1  & $[5; 6]$ & 0.91$\pm$0.02$\pm$0.08$\pm$0.03$\pm$0.06 & 0.156$\pm$0.011$\pm$0.005$\pm$0.008\\
 $[6; 7]$ & 14.9$\pm$0.4$\pm$0.8$\pm$0.6  & $[6; 8]$ & 0.98$\pm$0.02$\pm$0.09$\pm$0.05$\pm$0.06 & 0.057$\pm$0.005$\pm$0.003$\pm$0.003\\
 $[7; 8]$ & 8.3$\pm$0.3$\pm$0.5$\pm$0.3 & & &  \\
 $[8; 10]$ & 3.7$\pm$0.1$\pm$0.2$\pm$0.1 & & &  \\
 $[10; 15]$ & 0.77$\pm$0.03$\pm$0.05$\pm$0.03 & & &  \\
\end{tabular}										\caption{\label{tab:2}
Summary of the results on the inclusive \jpsi\ differential cross 
sections and nuclear modification factors for \mbox{p--Pb} collisions. The results of the cross section 
interpolation for pp collisions are also shown.
For p--Pb cross section results, the first quoted uncertainty is statistical. The following uncertainties 
are systematic, the second one being $p_{\rm T}$-uncorrelated and the third one $p_{\rm T}$-correlated.
For $R_{\rm pPb}$ the first quoted uncertainty is statistical. The following uncertainties are systematic, the second one being $p_{\rm T}$-uncorrelated. For dielectron results the third uncertainty is $p_{\rm T}$-correlated, while for dimuon results the third uncertainty is partially $p_{\rm T}$-correlated and the fourth is $p_{\rm T}$-correlated.
For the results on the interpolated \mbox{pp} cross section, the first quoted uncertainty combines statistical and $p_{\rm T}$-uncorrelated systematic uncertainties. For dielectron results the second uncertainty is $p_{\rm T}$-correlated systematic, while for dimuon results the second uncertainty is partially $p_{\rm T}$-correlated, and the third is $p_{\rm T}$-correlated.
}
\end{table}	
									
For the dielectron analysis, the $p_{\rm T}$-integrated cross section  was also determined, obtaining  
$${\rm d}\sigma^{{\rm J}/\psi}_{\rm pPb}/{\rm d}y (-1.37<y_{\rm cms}<0.43)= 909 \pm 78 ({\rm stat.}) \pm 71 ({\rm syst.}) \mu{\rm b}.$$ The corresponding \pt-integrated cross sections for the dimuon analysis were published in~\cite{Abelev:2013yxa}.

Starting from the \pt-differential \jpsi\ cross sections it is possible to evaluate, as additional information, the mean $p_{\rm T}$ ($\langle p_{{\mathrm T}} \rangle$) for the various $y$-ranges, by means of fits based on the empirical function:
\begin{equation}
\frac{{\rm d}^{2}\sigma^{\rm J/\psi}_{\rm pPb}}{{\rm d}y{\rm d}p_{\rm T}} = C \times \frac{p_{\rm T}}{\left[1+\left( \frac{p_{\rm T}}{p_{0}}\right)^{2}\right]^{n}}
\label{eq:pt}
\end{equation}
where $C$, $p_{0}$ and $n$ are free parameters.
The quality of the fits is satisfactory ($\chi^2/{\rm ndf}\sim 1$) and the resulting $\langle p_{{\mathrm T}} \rangle$ values, computed for the measured \pt\ ranges, are 
\begin{eqnarray*}
\langle p_{{\mathrm T}} \rangle (-4.46<y_{\rm cms}<-2.96) = 2.47 \pm 0.01 {\rm(stat.)} \pm 0.03 {\rm (syst.)}\, {\rm GeV}/c\\
\langle p_{{\mathrm T}} \rangle (-1.37<y_{\rm cms}<0.43) = 2.86 \pm 0.15 {\rm(stat.)} \pm 0.10 {\rm (syst.)}\, {\rm GeV}/c\\
\langle p_{{\mathrm T}} \rangle (2.03<y_{\rm cms}<3.53) = 2.77 \pm 0.01 {\rm(stat.)} \pm 0.03 {\rm (syst.)}\, {\rm GeV}/c \\
\end{eqnarray*}
The quoted uncertainties were obtained by performing fits including only statistical (or uncorrelated systematic) uncertainties on differential cross sections.

In order to perform a meaningful comparison of $\langle p_{{\mathrm T}} \rangle$ results in the dielectron and dimuon analysis, the values from the dimuon analysis have also been extracted, with the same procedure detailed above, in the range $p_{\rm T}<10$ GeV/$c$, obtaining results which are smaller by less than 2\% with respect to the full \pt\ range. It is found that $\langle p_{{\mathrm T}} \rangle$ is larger at central rapidity.
Furthermore, the $\langle p_{{\mathrm T}} \rangle$ measured at forward $y_{\rm cms}$ is significantly larger than at backward $y_{\rm cms}$. This difference, which could be partly due to the slightly different $|y|$-coverage, persists when $\langle p_{{\mathrm T}} \rangle$ is calculated in the $|y_{\rm cms}|$ region common to \mbox{p--Pb} and \mbox{Pb--p} ($2.96 < |y_{\rm cms}| < 3.53$). The values obtained in this case are $2.58 \pm 0.02 {\rm(stat.)} \pm 0.04 {\rm (syst.)}\, {\rm GeV}/c$ and $2.69 \pm 0.02 {\rm(stat.)} \pm 0.03 {\rm (syst.)}\, {\rm GeV}/c$, respectively at backward and forward $y_{\rm cms}$, and differ by $\sim 2\sigma$.

The \jpsi\ nuclear modification factor \RpPb\ is obtained as the ratio of the differential cross sections between proton-nucleus and proton-proton collisions,
 normalized to $A_{\rm Pb}$:
\begin{equation}
R_{\rm pPb}(y,p_{\rm T})=\frac{{\rm d}^{2}\sigma^{\rm J/\psi}_{\rm pPb}/{\rm d}y{\rm d}p_{\rm T}}{A_{\rm Pb} \cdot
{\rm d}^{2}\sigma^{\rm J/\psi}_{\rm pp}/{\rm d}y{\rm d}p_{\rm T}}
\end{equation}

Since no \mbox{pp} data are available at \sqrtspp, the ${\rm d}^{2}\sigma^{\rm J/\psi}_{\rm pp}/{\rm d}y{\rm d}p_{\rm
T}$ reference cross sections were obtained by means of an interpolation/extrapolation procedure. 
For the dielectron analysis, the starting point of the interpolation procedure is the determination of $\textrm{d} \sigma / \textrm{d} y$ for inclusive \jpsi\ in pp collisions at $y_{\rm cms} \sim 0$ and \sqrtspp, carried out as for the analysis described in~\cite{Abelev:2013ila}. Available mid-rapidity data at $\sqrt{s}$ = 0.2\cite{Adare:2006kf}, 1.96\cite{Acosta:2004yw}, 2.76\cite{Abelev:2012kr} and 7 TeV\cite{Aamodt:2011gj} are interpolated using several empirical functions (exponential, logarithmic and power-law, covering in this way the various possibilities for the curvature of the $\sqrt{s}$-dependence) obtaining $\textrm{d} \sigma / \textrm{d} y = 6.19 \pm 1.03$  $\mu$b. Even if the $y_{\rm cms}$ range covered in this analysis is shifted by 0.465 units with respect to mid-rapidity,  the rapidity-dependence of the cross section is negligible compared to the uncertainty on the interpolation procedure.
Then, a method similar to the one in~\cite{Bossu:2011qe} is applied to derive the \pt-differential cross section. It is based on the empirical observation that
\mbox{pp} and p$\overline{\textrm{p}}$ results  on differential spectra obtained at various collision energies and in different rapidity ranges~\cite{Aamodt:2011gj,Acosta:2004yw,Adare:2006kf,Aaij:2013yaa,Aaij:2011jh} exhibit 
scaling properties when plotted as a function of \pt/$\langle p_{{\mathrm T}} \rangle$. The normalized spectra, with the statistical and the bin-by-bin uncorrelated systematic uncertainties added in quadrature, can be fitted with a one-parameter function described in~\cite{Bossu:2011qe}. The \pt-differential cross sections at mid-rapidity and \sqrtspp\ can then be obtained by rescaling the fitted universal distribution using the previously estimated $\textrm{d} \sigma / \textrm{d} y$ and its corresponding $\langle p_{{\mathrm T}} \rangle$. The latter value is obtained by an interpolation of the energy-dependence of $\langle p_{{\mathrm T}} \rangle$ values evaluated fitting the available experimental mid-rapidity results \cite{Aamodt:2011gj,Acosta:2004yw,Adare:2006kf} with  exponential, logarithmic and power-law functions. One obtains in this way, in the range $p_{\rm T}<10$ GeV/$c$,  $\langle p_{{\mathrm T}} \rangle = 2.81 \pm 0.10$ GeV/$c$ as an average of the results calculated with the various empirical functions.
As outlined above for $\textrm{d} \sigma / \textrm{d} y$, the 0.465 $y$-unit shift of the data with respect to mid-rapidity has a negligible effect also on $\langle p_{{\mathrm T}} \rangle$.

For the dimuon analysis, thanks to the smaller uncertainties with respect to mid-rapidity results, an approach equivalent to that described in~\cite{ALICE:2013spa}, exclusively based on the ALICE data collected at $\sqrt{s} = 2.76$ TeV~\cite{Abelev:2012kr} and 7 TeV~\cite{Abelev:2014qha} in $2.5<y_{\rm cms}<4$, $p_{\rm T}<8$ GeV/$c$ has been used. The reference cross sections are obtained with a two-step
procedure, corresponding to an energy interpolation followed by a rapidity extrapolation. In the first step, for each \pt\ bin, the 
${\rm d}^{2}\sigma^{\rm J/\psi}_{\rm pp}/{\rm d}y{\rm d}p_{\rm T}$ values at 
$\sqrt{s} = 2.76$ and 7 TeV are interpolated, using three different empirical functions (linear, power-law and exponential) 
to estimate the cross section values at \sqrtspp.
The central values are calculated as the average of the 
results obtained with the three functions, while the associated uncertainties come from the experimental uncertainties on the points used for the interpolation, added in quadrature to a contribution chosen as the maximum spread of the results from the different interpolating functions.
In the second step, this result is extrapolated from $2.5<y
_{\rm cms}<4$ to the \mbox{p--Pb} and \mbox{Pb--p} $y_{\rm cms}$ ranges, using the  scaling factors for the \pt-integrated cross sections computed in~\cite{ALICE:2013spa}.
Finally, since the LHCb Collaboration has shown that the \jpsi\ \pt\ distributions slightly depend on $y_{\rm cms}$~\cite{Aaij:2011jh} in the rapidity range covered in the dimuon analysis, a \pt-dependent correction tuned on these data (10\% maximum at large \pt) is applied.

\begin{figure}[htbp]
\centering
\resizebox{0.6\textwidth}{!}
{\includegraphics{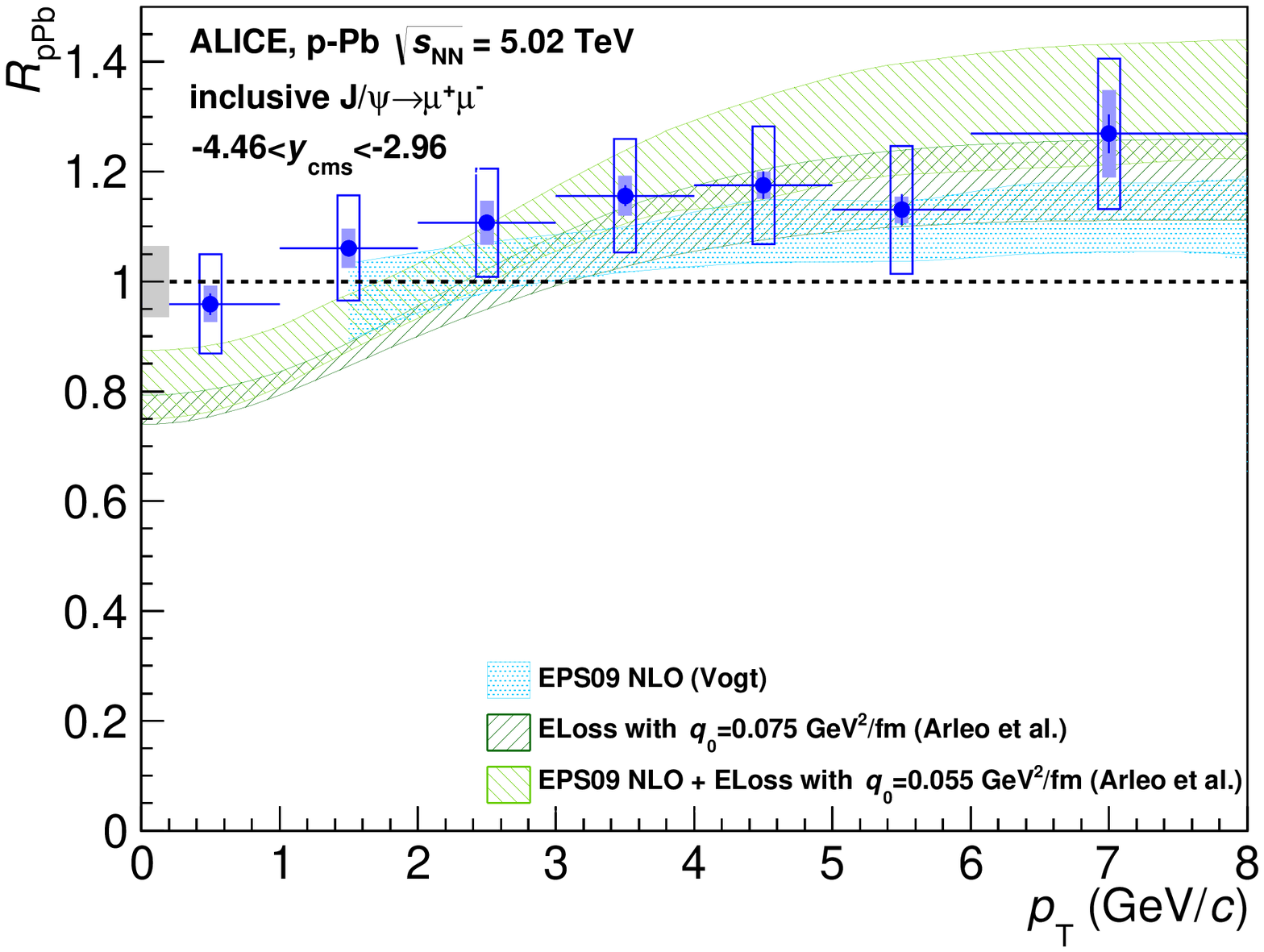}}
\resizebox{0.6\textwidth}{!}
{\includegraphics{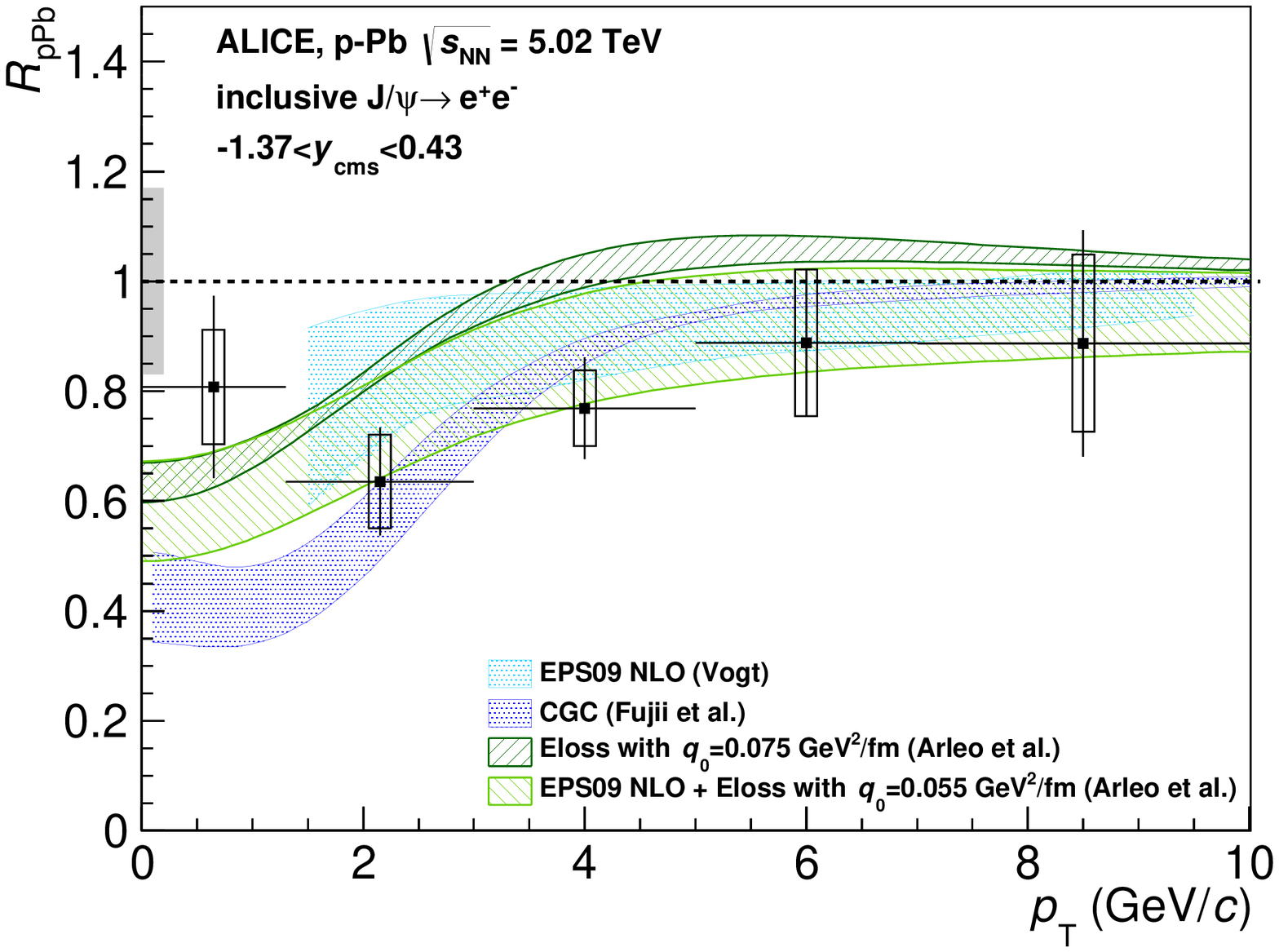}}
\resizebox{0.6\textwidth}{!}
{\includegraphics{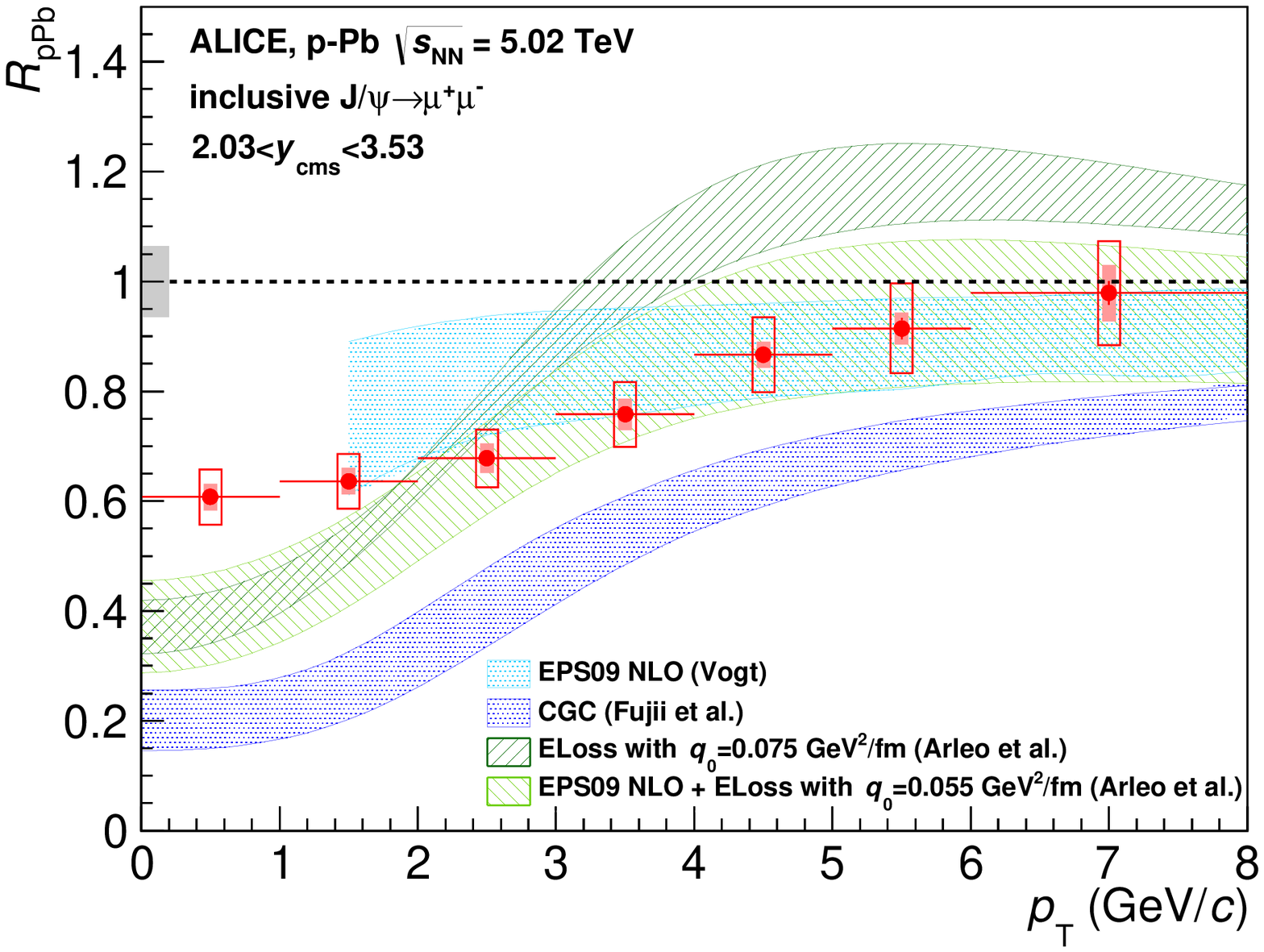}}
\caption{The \jpsi\ nuclear modification factor as a function of \pt\ at backward (top), mid (center) and forward (bottom) rapidities. 
Statistical uncertainties are represented by vertical error bars, while open boxes correspond to uncorrelated uncertainties and the shaded areas to uncertainties 
partially correlated in \pt. The boxes around $R_{\rm pPb}=1$ show the size of the correlated uncertainties. 
The horizontal bars correspond to the widths of the \pt\ bins. Results from various models are also shown, including a pure shadowing calculation ~\cite{Albacete:2013ei} based on the EPS09 parameterization, a CGC-inspired model~\cite{Fujii:2013gxa}, and the results of the coherent energy loss calculation~\cite{Arleo:2013zua}, with or without the inclusion of an EPS09 shadowing contribution.}
\label{fig:RpA}
\end{figure}

The inclusive \jpsi\ nuclear modification factor is shown in Fig.~\ref{fig:RpA} for the three rapidity regions under study. The numerical values of $R_{\rm pPb}$, as well as the results of the interpolation procedure for the estimate of the pp cross sections, can be found in Table~\ref{tab:2}. For the dimuon analysis, the evaluation of $R_{\rm pPb}$ is restricted to $p_{\rm T}< 8$ GeV/$c$, 
the region covered by the \mbox{pp} measurements used in the evaluation of the reference cross sections.
The sources of systematic uncertainties on $R_{\rm pPb}$ and their values are summarized in Table~\ref{tab:syst}. The terms related to the \mbox{pp} reference cross sections contribute to uncorrelated, partially or fully correlated uncertainties on $R_{\rm pPb}$, depending on their origin. In particular, for the dimuon analysis: (i) the statistical and \pt-uncorrelated systematic uncertainties on the $\sqrt{s}=2.76$ and 7 TeV \mbox{pp} data contribute to the uncorrelated uncertainty; (ii) the spread of the results obtained with various interpolating/extrapolating functions in $\sqrt{s}$ and $y_{\rm cms}$ contribute to the partially correlated uncertainty; (iii) the $\sqrt{s}$-correlated uncertainties between the $\sqrt{s}=2.76$ and 7 TeV \mbox{pp} data contribute to the correlated uncertainty. At forward and mid-rapidity the \jpsi\ \RpPb\ shows a clear suppression at low \pt, vanishing at high \pt. At backward rapidity no suppression is present, within uncertainties. 

For the dielectron analysis, the $p_{\rm T}$-integrated nuclear modification factor was also calculated, carrying out the signal extraction procedure on the $p_{\rm T}$-integrated invariant mass spectrum. The obtained value 
$$R_{\rm pPb}=0.71 \pm 0.06 ({\rm stat.}) \pm 0.13 ({\rm syst.})$$ 
is consistent with the forward rapidity ($2.03 < y_{cms} <3.53$) dimuon result,  and smaller than the backward one ($-4.46 < y_{cms} <-2.96$) by $\sim 2\sigma$ ~\cite{Abelev:2013yxa}.

In Fig.~\ref{fig:RpA} predictions from various models are compared to the data. 
A calculation based on the next-to-leading order (NLO) Color Evaporation Model (CEM) for the prompt \jpsi\ production and the EPS09 
shadowing parametrization \cite{Albacete:2013ei} reproduces within uncertainties the \pt-dependence and the amplitude of the suppression for
\pt\ $>1.5$ GeV/$c$ in the three rapidity regions under study. The theoretical uncertainties arise from the uncertainties on EPS09 as well as on the values of charm quark mass and of the renormalization and factorization scales used for the cross section calculation.
Data are also compared  to two calculations based on a parametrization of experimental results on prompt \jpsi\ production in \mbox{pp} collisions and including the effects of coherent energy loss \cite{Arleo:2013zua} in the cold nuclear medium. One of the calculations includes only coherent energy loss, while the other combines coherent energy loss with EPS09 shadowing. The uncertainty bands include, for the coherent energy loss mechanism, a variation of both the $q_{\rm 0}$ parameter (gluon transport coefficient evaluated at $x = 0.01$) and the parametrization of the production cross section. At forward rapidity the pure energy loss scenario predicts a much steeper \pt-dependence, while better agreement is found when the EPS09 contribution is included. However, at low \pt, a discrepancy between data and both calculations is observed.
Also at mid-rapidity the coherent energy loss model including the EPS09 contribution better describes the data, although the larger uncertainties prevent a firm conclusion. The same features can be observed at backward rapidity, where the calculation including coherent energy loss and shadowing agrees with the data in showing weak nuclear effects on \jpsi\ production.
Finally, the results at central and forward rapidities are compared with a prediction based on the CGC framework and using CEM for the prompt \jpsi\ production \cite{Fujii:2013gxa}. 
In the backward rapidity region, higher gluon $x$ in the nucleus are probed and the CGC model is out of its range of applicability.
The quoted uncertainties are related to the choices of $Q^2_s$ and of the charm quark mass. While the model is in fair agreement with mid-rapidity data, it clearly underpredicts the \jpsi\ $R_{\rm pPb}$ in the full \pt\ range at forward rapidity. 

The theoretical calculations discussed above are carried out for prompt \jpsi\  (i.e., direct \jpsi\ and the contribution from $\chi_{c}$ and $\psi$(2S) decays), while the measurements are for inclusive \jpsi\ which include a non-prompt contribution from B-hadron decays. The contribution of the latter source to $R^{\rm incl}_{\rm pPb}$ can be evaluated from the measured fraction $f_{B}$ of non-prompt to prompt \jpsi\ production in \mbox{pp} collisions and on the suppression $R^{\rm non-prompt}_{\rm pPb}$ of non-prompt \jpsi\ in \mbox{p--Pb} collisions. More in detail,
in the range $2 <y_{\rm cms}< 4.5$, the fraction $f_{B}$ measured by LHCb in \mbox{pp} collisions at $\sqrt{s} = 7$ TeV, increases  from 0.08 to 0.22 from $p_{\rm T}=0$ to 8 GeV/$c$~\cite{Aaij:2011jh}. This quantity has a small variation within the $y_{\rm cms}$ range covered and is also not strongly $\sqrt{s}$-dependent (similar values are obtained for $\sqrt{s} = 8$ TeV~\cite{Aaij:2013yaa}). At mid-rapidity, $f_{\rm B}$ was measured by ALICE in pp collisions at $\sqrt{s}=7$ TeV and ranges from 0.10 to 0.44 for \pt\ increasing from 1.3 to 10 GeV/$c$~\cite{Abelev:2012gx}.  
$R^{\rm non-prompt}_{\rm pPb}$ was measured at \sqrts\ by LHCb, integrated over \pt, obtaining 0.83 $\pm$ 0.02 $\pm$ 0.08 for $2.5 < y_{\rm cms} < 4$ and 0.98 $\pm$ 0.06 $\pm$ 0.10 for $-4 <y_{\rm cms}< -2.5$~\cite{Aaij:2013zxa}.
Assuming for each \pt-bin a variation of $R^{\rm non-prompt}_{\rm pPb}$ between 0.6 and 1.3, a conservative choice due to the unavailability of a \pt-differential result,  and considering the \pt-dependence of $f_{B}$ at $\sqrt{s}=7$ TeV, one can extract 
$R^{\rm prompt}_{\rm pPb}$ as $R^{\rm prompt}_{\rm pPb}=R^{\rm incl}_{\rm pPb}+f_{\rm B}\cdot(R^{\rm incl}_{\rm pPb}-R^{\rm non-prompt}_{\rm pPb})$.
The maximum differences between the inclusive and prompt \RpPb\ obtained in this way are, for low and high \pt: (i) 3 and 10\% at backward rapidity; (ii) 11 and 16\% at central rapidity; (iii) 10 and 8\% at forward rapidity. These variations are, at most, of the same order of magnitude as the quoted uncertainties on inclusive \RpPb.

The \RpPb\ results shown in this paper can be considered as a valuable tool to improve our understanding of the contribution of CNM to the suppression of the \jpsi\ yields observed in \mbox{Pb--Pb}~\cite{Abelev:2013ila,Adam:2015rba}. Indeed, as verified in~\cite{Abelev:2013yxa} for the dimuon analysis, in \mbox{Pb--Pb} collisions the Bjorken-$x$ ranges probed by the \jpsi\ production process in the two colliding nuclei, assuming a $gg\rightarrow J/\psi$ (2$\rightarrow$1)~\cite{Vogt:2004dh} mechanism, are shifted by only $\sim$10\% with respect to the corresponding intervals for \mbox{p--Pb} and \mbox{Pb--p}, despite the different energy ($\sqrt{s_{\rm NN}}$ = 2.76 TeV) and the slightly different $y_{\rm cms}$ range ($2.5 < y < 4$) for \mbox{Pb--Pb}. A similar conclusion holds at mid-rapidity, where the covered $x$-intervals, calculated for $p_{\rm T}=\langle p_{\rm T}\rangle$, are $6.1\times 10^{-4} < x < 3.0\times 10^{-3}$ and $7.0\times 10^{-4} < x < 3.5\times 10^{-3}$ for \mbox{p--Pb} and 
\mbox{Pb--Pb} collisions, respectively.
Under the assumption that shadowing is the main CNM-related mechanism that plays a role in the \jpsi\ production and that its effect on the two colliding nuclei in \mbox{Pb--Pb} collisions can be factorized, the product $R_{\mathrm{pPb}} \times R_{\mathrm{Pbp}}$ ($R^{2}_{\mathrm{pPb}}$) can be considered as an estimate of CNM effects in \mbox{Pb--Pb} collisions at forward (central) rapidity~\cite{Vogt:2010aa,Ferreiro:2008wc}. This conclusion holds not only for the $2\rightarrow 1$ production process but also when the more general $2\rightarrow 2$ mechanism ($gg\rightarrow J/\psi g$) is considered.

\begin{figure}[htbp]
\centering
\resizebox{0.7\textwidth}{!}
{\includegraphics{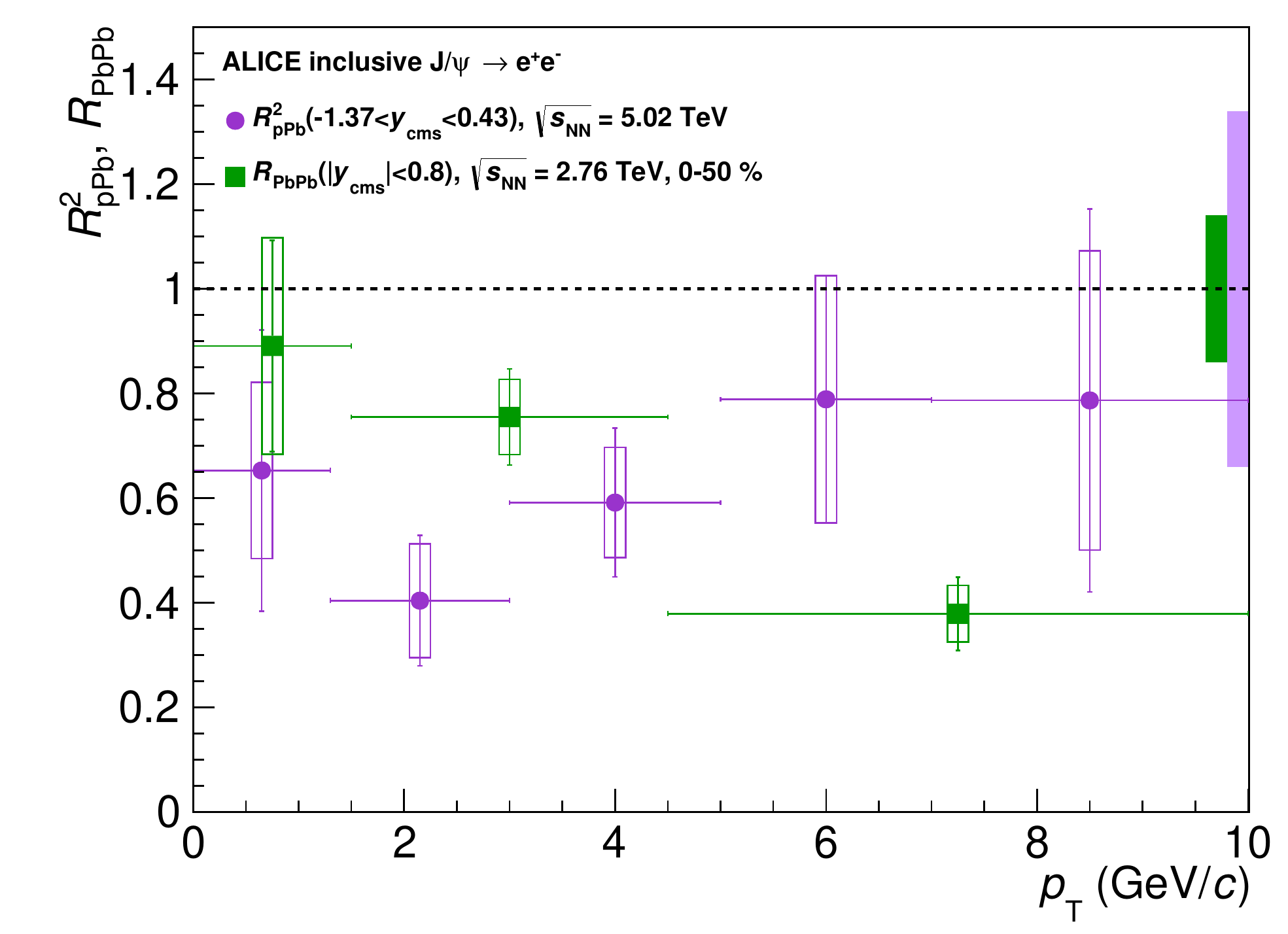}}
\resizebox{0.7\textwidth}{!}
{\includegraphics{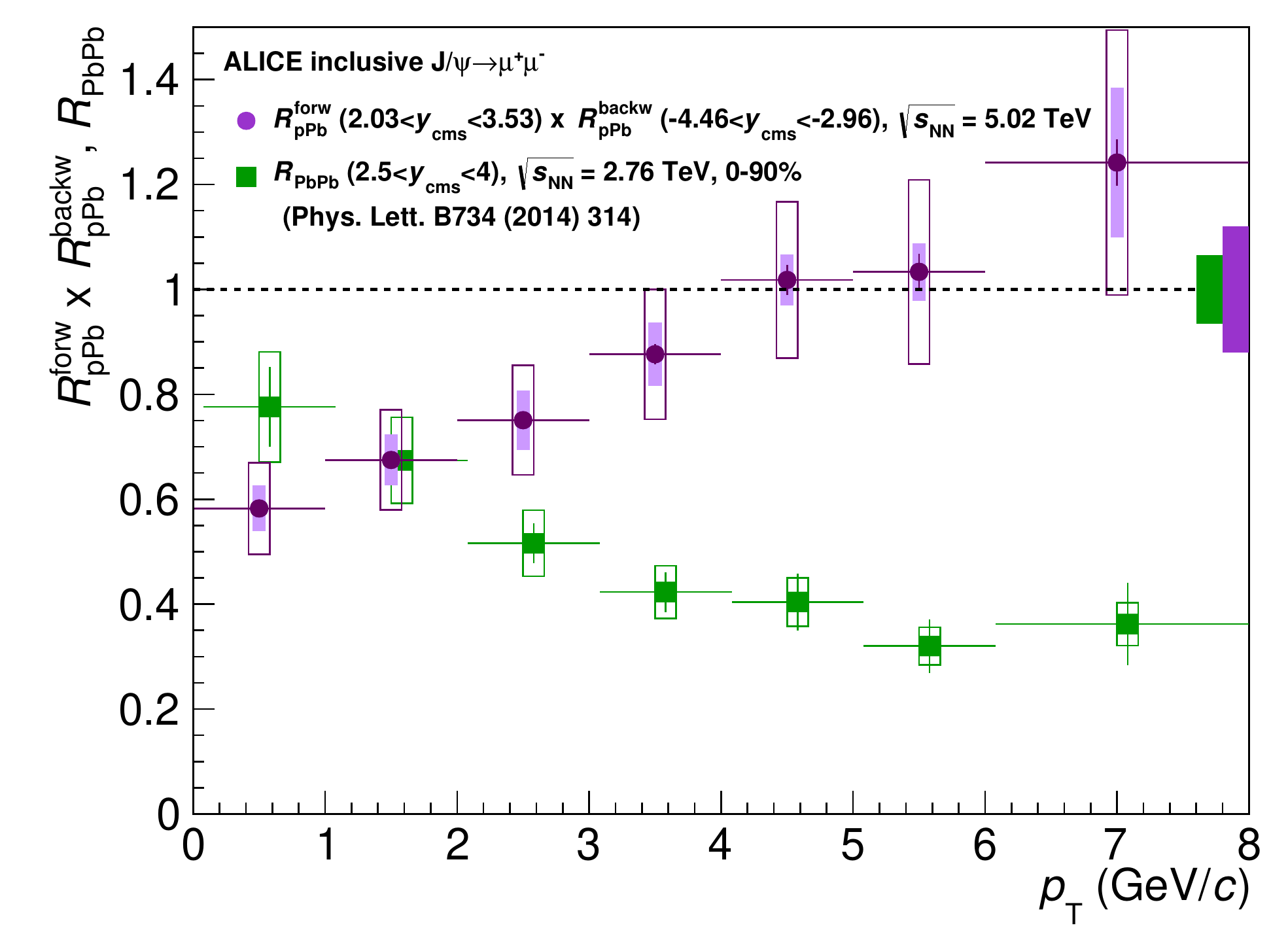}}
\caption{ The estimate of the \pt-dependence of CNM effects in \mbox{Pb--Pb}, calculated as $R^2_{\mathrm{pPb}}$ for mid-rapidity data (top) and as $R_{\mathrm{pPb}} \times R_{\mathrm{Pbp}}$ (bottom) at forward rapidity.  The quantities are compared to $R_{\mathrm{PbPb}}$ measured in
\mbox{Pb--Pb} collisions in the (approximately) corresponding $y$-ranges~\cite{Abelev:2013ila,Adam:2015rba}. 
The vertical error bars correspond to the statistical uncertainties, the open boxes (shaded areas) represent $p_{\rm T}$-uncorrelated (partially correlated) 
systematic uncertainties, while the boxes around $R_{\rm pPb}=1$ show the size of the correlated uncertainties. The horizontal bars correspond to the widths of the \pt\ bins. The \mbox{Pb--Pb} points in the bottom panel were slightly displaced in \pt, to improve visibility.}
\label{fig:RpARAp}
\end{figure}

In Fig.~\ref{fig:RpARAp} the comparison of the measured $R_{\rm PbPb}$ with the quantities defined above is carried out. Such a comparison should be considered as qualitative, in view of the slight $x$-mismatch detailed above and of the fact that, at mid-rapidity, the centrality ranges probed in \mbox{p--Pb} and \mbox{Pb--Pb} are not the same (0-100\% and 0-50\%, respectively).
In both rapidity regions, the extrapolation of CNM effects shows a clear \pt-dependence, corresponding to a strong suppression at low \pt, which vanishes for large transverse momenta.  
At low \pt\ and central rapidity, there might be an indication for a \mbox{Pb--Pb} suppression smaller than the CNM extrapolation, consistent with 
the presence of a contribution related to the (re)combination of $c \bar{c}$ pairs~\cite{Abelev:2013ila}, taking place in the hot medium. A similar effect can be seen at forward rapidity. At large \pt\ and forward rapidity, the observed suppression in \mbox{Pb--Pb} collisions is much larger than CNM extrapolations, showing that, in this transverse-momentum region, suppression effects in hot matter, possibly related to color screening, become dominant. 
   
\begin{figure}[htbp]
\centering
\resizebox{0.7\textwidth}{!}
{\includegraphics{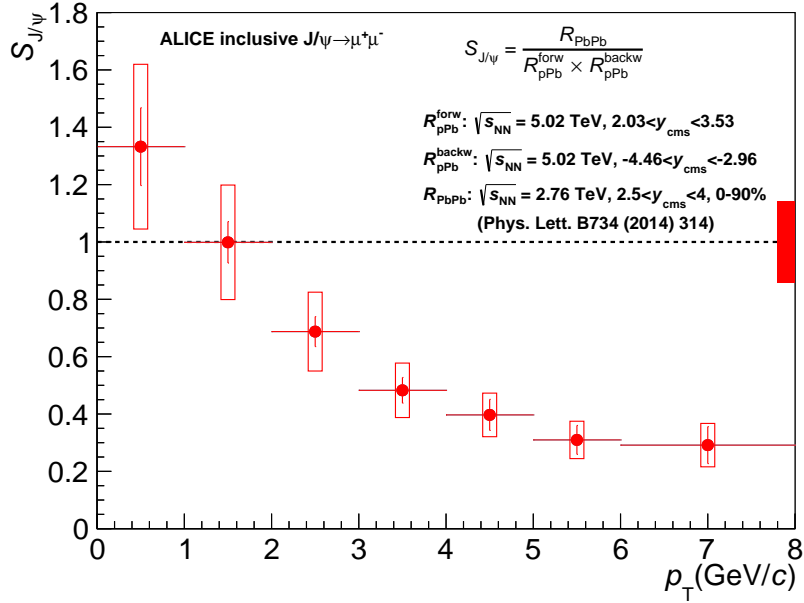}}
\caption{The ratio between $R_{\mathrm{PbPb}}$ for inclusive \jpsi\  at forward rapidity and the product $R_{\mathrm{pPb}} \times R_{\mathrm{Pbp}}$ of the nuclear modification factors at forward and backward rapidity. None of the uncertainties cancels out in the ratio. Statistical uncertainties are shown as vertical error bars, while the boxes around the points represent a quadratic combination of uncorrelated and partially correlated systematic uncertainties. The box around $S_{\rm J/\psi}=1$ corresponds to correlated uncertainties. 
The horizontal bars coincide with the widths of the \pt\ bins.}
\label{fig:MeasExp}
\end{figure}

Finally, a more direct comparison of \mbox{Pb--Pb} results with the CNM extrapolation can be obtained by defining the ratio $S_{\rm J/\psi} = R_{\mathrm{PbPb}}/ (R_{\mathrm{pPb}} \times R_{\mathrm{Pbp}})$. Such a quantity, for forward rapidity results, is shown in Fig.~\ref{fig:MeasExp} and confirms the main features detailed above, i.e., a strong suppression of \jpsi\ at large \pt, and a hint for an enhancement at low \pt. At central rapidity, due to the sizeable uncertainties on both \mbox{p--Pb} and \mbox{Pb--Pb} results, only the $p_{\rm T}$-integrated ratio can be obtained. Using the $R_{\rm PbPb}$ in the 0-90\% centrality range~\cite{Abelev:2013ila}, and the integrated $R_{\rm pPb}$ given above, one gets $1.43 \pm 0.26 {\rm(stat)} \pm 0.56 {\rm(syst)}$. More precise measurements are needed to draw a firm conclusion in this rapidity range. 

In summary, we have presented results on the inclusive \jpsi\ production in \mbox{p--Pb} collisions at \sqrts. The \pt-differential  cross sections, the $\langle p_{\rm T}\rangle$ and the nuclear modification factors have been evaluated in three rapidity regions: $-4.46 <y_{\rm cms}< -2.96$, $-1.37 < y_{\rm cms} < 0.43$ and $2.03 <y_{\rm cms} < 3.53$. 
At forward and mid-rapidity a significant suppression is observed at low \pt, with a vanishing trend at high \pt. At backward rapidity no significant suppression or enhancement is visible. Comparisons with theoretical models based on a combination of nuclear shadowing and coherent energy loss effects provide a fair description of the observed patterns, except at forward rapidity and low transverse momentum.
These results can be used to provide a qualitative estimate of the influence of cold nuclear matter effects on the \jpsi\ suppression observed in \mbox{Pb--Pb} collisions. Under the assumption that shadowing represents the main CNM  contribution, we find that it cannot account for the observed suppression in \mbox{Pb--Pb} at high \pt. At low \pt, the observed CNM effects alone may suggest a suppression larger than that observed in \mbox{Pb--Pb}, which is consistent with the presence of a charm quark (re)combination component to the \jpsi\ production in nucleus-nucleus collisions.

%
\newenvironment{acknowledgement}{\relax}{\relax}
\begin{acknowledgement}
\section*{Acknowledgements}
The ALICE Collaboration would like to thank all its engineers and technicians for their invaluable contributions to the construction of the experiment and the CERN accelerator teams for the outstanding performance of the LHC complex.
The ALICE Collaboration gratefully acknowledges the resources and support provided by all Grid centres and the Worldwide LHC Computing Grid (WLCG) collaboration.
The ALICE Collaboration acknowledges the following funding agencies for their support in building and
running the ALICE detector:
State Committee of Science,  World Federation of Scientists (WFS)
and Swiss Fonds Kidagan, Armenia,
Conselho Nacional de Desenvolvimento Cient\'{\i}fico e Tecnol\'{o}gico (CNPq), Financiadora de Estudos e Projetos (FINEP),
Funda\c{c}\~{a}o de Amparo \`{a} Pesquisa do Estado de S\~{a}o Paulo (FAPESP);
National Natural Science Foundation of China (NSFC), the Chinese Ministry of Education (CMOE)
and the Ministry of Science and Technology of China (MSTC);
Ministry of Education and Youth of the Czech Republic;
Danish Natural Science Research Council, the Carlsberg Foundation and the Danish National Research Foundation;
The European Research Council under the European Community's Seventh Framework Programme;
Helsinki Institute of Physics and the Academy of Finland;
French CNRS-IN2P3, the `Region Pays de Loire', `Region Alsace', `Region Auvergne' and CEA, France;
German Bundesministerium fur Bildung, Wissenschaft, Forschung und Technologie (BMBF) and the Helmholtz Association;
General Secretariat for Research and Technology, Ministry of
Development, Greece;
Hungarian Orszagos Tudomanyos Kutatasi Alappgrammok (OTKA) and National Office for Research and Technology (NKTH);
Department of Atomic Energy and Department of Science and Technology of the Government of India;
Istituto Nazionale di Fisica Nucleare (INFN) and Centro Fermi -
Museo Storico della Fisica e Centro Studi e Ricerche "Enrico
Fermi", Italy;
MEXT Grant-in-Aid for Specially Promoted Research, Ja\-pan;
Joint Institute for Nuclear Research, Dubna;
National Research Foundation of Korea (NRF);
Consejo Nacional de Cienca y Tecnologia (CONACYT), Direccion General de Asuntos del Personal Academico(DGAPA), M\'{e}xico, :Amerique Latine Formation academique – European Commission(ALFA-EC) and the EPLANET Program
(European Particle Physics Latin American Network)
Stichting voor Fundamenteel Onderzoek der Materie (FOM) and the Nederlandse Organisatie voor Wetenschappelijk Onderzoek (NWO), Netherlands;
Research Council of Norway (NFR);
National Science Centre, Poland;
Ministry of National Education/Institute for Atomic Physics and Consiliul Naţional al Cercetării Ştiinţifice - Executive Agency for Higher Education Research Development and Innovation Funding (CNCS-UEFISCDI) - Romania;
Ministry of Education and Science of Russian Federation, Russian
Academy of Sciences, Russian Federal Agency of Atomic Energy,
Russian Federal Agency for Science and Innovations and The Russian
Foundation for Basic Research;
Ministry of Education of Slovakia;
Department of Science and Technology, South Africa;
Centro de Investigaciones Energeticas, Medioambientales y Tecnologicas (CIEMAT), E-Infrastructure shared between Europe and Latin America (EELA), Ministerio de Econom\'{i}a y Competitividad (MINECO) of Spain, Xunta de Galicia (Conseller\'{\i}a de Educaci\'{o}n),
Centro de Aplicaciones Tecnológicas y Desarrollo Nuclear (CEA\-DEN), Cubaenerg\'{\i}a, Cuba, and IAEA (International Atomic Energy Agency);
Swedish Research Council (VR) and Knut $\&$ Alice Wallenberg
Foundation (KAW);
Ukraine Ministry of Education and Science;
United Kingdom Science and Technology Facilities Council (STFC);
The United States Department of Energy, the United States National
Science Foundation, the State of Texas, and the State of Ohio;
Ministry of Science, Education and Sports of Croatia and  Unity through Knowledge Fund, Croatia.
Council of Scientific and Industrial Research (CSIR), New Delhi, India
\end{acknowledgement}
\bibliographystyle{utphys}
\bibliography{biblio.bib}
\newpage
\appendix
\section{The ALICE Collaboration}



\begingroup
\small
\begin{flushleft}
J.~Adam\Irefn{org39}\And
D.~Adamov\'{a}\Irefn{org82}\And
M.M.~Aggarwal\Irefn{org86}\And
G.~Aglieri Rinella\Irefn{org36}\And
M.~Agnello\Irefn{org110}\And
N.~Agrawal\Irefn{org47}\And
Z.~Ahammed\Irefn{org130}\And
I.~Ahmed\Irefn{org16}\And
S.U.~Ahn\Irefn{org67}\And
I.~Aimo\Irefn{org93}\textsuperscript{,}\Irefn{org110}
% ...
\renewcommand\labelenumi{\textsuperscript{\theenumi}~}

\section*{Affiliation notes}
\renewcommand\theenumi{\roman{enumi}}
\begin{Authlist}
\item \Adef{0}Deceased
\end{Authlist}

\section*{Collaboration Institutes}
\renewcommand\theenumi{\arabic{enumi}~}
\begin{Authlist}

\item \Idef{org1}A.I. Alikhanyan National Science Laboratory (Yerevan Physics Institute) Foundation, Yerevan, Armenia
\item \Idef{org2}Benem\'{e}rita Universidad Aut\'{o}noma de Puebla, Puebla, Mexico
\item \Idef{org3}Bogolyubov Institute for Theoretical Physics, Kiev, Ukraine
\item \Idef{org4}Bose Institute, Department of Physics and Centre for Astroparticle Physics and Space Science (CAPSS), Kolkata, India
\item \Idef{org5}Budker Institute for Nuclear Physics, Novosibirsk, Russia
\item \Idef{org6}California Polytechnic State University, San Luis Obispo, California, United States
\item \Idef{org7}Central China Normal University, Wuhan, China
\item \Idef{org8}Centre de Calcul de l'IN2P3, Villeurbanne, France
\item \Idef{org9}Centro de Aplicaciones Tecnol\'{o}gicas y Desarrollo Nuclear (CEADEN), Havana, Cuba
\item \Idef{org10}Centro de Investigaciones Energ\'{e}ticas Medioambientales y Tecnol\'{o}gicas (CIEMAT), Madrid, Spain
\item \Idef{org11}Centro de Investigaci\'{o}n y de Estudios Avanzados (CINVESTAV), Mexico City and M\'{e}rida, Mexico
\item \Idef{org12}Centro Fermi - Museo Storico della Fisica e Centro Studi e Ricerche ``Enrico Fermi'', Rome, Italy
\item \Idef{org13}Chicago State University, Chicago, Illinois, USA
\item \Idef{org14}China Institute of Atomic Energy, Beijing, China
\item \Idef{org15}Commissariat \`{a} l'Energie Atomique, IRFU, Saclay, France
\item \Idef{org16}COMSATS Institute of Information Technology (CIIT), Islamabad, Pakistan
\item \Idef{org17}Departamento de F\'{\i}sica de Part\'{\i}culas and IGFAE, Universidad de Santiago de Compostela, Santiago de Compostela, Spain
\item \Idef{org18}Department of Physics and Technology, University of Bergen, Bergen, Norway
\item \Idef{org19}Department of Physics, Aligarh Muslim University, Aligarh, India
\item \Idef{org20}Department of Physics, Ohio State University, Columbus, Ohio, United States
\item \Idef{org21}Department of Physics, Sejong University, Seoul, South Korea
\item \Idef{org22}Department of Physics, University of Oslo, Oslo, Norway
\item \Idef{org23}Dipartimento di Elettrotecnica ed Elettronica del Politecnico, Bari, Italy
\item \Idef{org24}Dipartimento di Fisica dell'Universit\`{a} 'La Sapienza' and Sezione INFN Rome, Italy
\item \Idef{org25}Dipartimento di Fisica dell'Universit\`{a} and Sezione INFN, Cagliari, Italy
\item \Idef{org26}Dipartimento di Fisica dell'Universit\`{a} and Sezione INFN, Trieste, Italy
\item \Idef{org27}Dipartimento di Fisica dell'Universit\`{a} and Sezione INFN, Turin, Italy
\item \Idef{org28}Dipartimento di Fisica e Astronomia dell'Universit\`{a} and Sezione INFN, Bologna, Italy
\item \Idef{org29}Dipartimento di Fisica e Astronomia dell'Universit\`{a} and Sezione INFN, Catania, Italy
\item \Idef{org30}Dipartimento di Fisica e Astronomia dell'Universit\`{a} and Sezione INFN, Padova, Italy
\item \Idef{org31}Dipartimento di Fisica `E.R.~Caianiello' dell'Universit\`{a} and Gruppo Collegato INFN, Salerno, Italy
\item \Idef{org32}Dipartimento di Scienze e Innovazione Tecnologica dell'Universit\`{a} del  Piemonte Orientale and Gruppo Collegato INFN, Alessandria, Italy
\item \Idef{org33}Dipartimento Interateneo di Fisica `M.~Merlin' and Sezione INFN, Bari, Italy
\item \Idef{org34}Division of Experimental High Energy Physics, University of Lund, Lund, Sweden
\item \Idef{org35}Eberhard Karls Universit\"{a}t T\"{u}bingen, T\"{u}bingen, Germany
\item \Idef{org36}European Organization for Nuclear Research (CERN), Geneva, Switzerland
\item \Idef{org37}Faculty of Engineering, Bergen University College, Bergen, Norway
\item \Idef{org38}Faculty of Mathematics, Physics and Informatics, Comenius University, Bratislava, Slovakia
\item \Idef{org39}Faculty of Nuclear Sciences and Physical Engineering, Czech Technical University in Prague, Prague, Czech Republic
\item \Idef{org40}Faculty of Science, P.J.~\v{S}af\'{a}rik University, Ko\v{s}ice, Slovakia
\item \Idef{org41}Faculty of Technology, Buskerud and Vestfold University College, Vestfold, Norway
\item \Idef{org42}Frankfurt Institute for Advanced Studies, Johann Wolfgang Goethe-Universit\"{a}t Frankfurt, Frankfurt, Germany
\item \Idef{org43}Gangneung-Wonju National University, Gangneung, South Korea
\item \Idef{org44}Gauhati University, Department of Physics, Guwahati, India
\item \Idef{org45}Helsinki Institute of Physics (HIP), Helsinki, Finland
\item \Idef{org46}Hiroshima University, Hiroshima, Japan
\item \Idef{org47}Indian Institute of Technology Bombay (IIT), Mumbai, India
\item \Idef{org48}Indian Institute of Technology Indore, Indore (IITI), India
\item \Idef{org49}Inha University, Incheon, South Korea
\item \Idef{org50}Institut de Physique Nucl\'eaire d'Orsay (IPNO), Universit\'e Paris-Sud, CNRS-IN2P3, Orsay, France
\item \Idef{org51}Institut f\"{u}r Informatik, Johann Wolfgang Goethe-Universit\"{a}t Frankfurt, Frankfurt, Germany
\item \Idef{org52}Institut f\"{u}r Kernphysik, Johann Wolfgang Goethe-Universit\"{a}t Frankfurt, Frankfurt, Germany
\item \Idef{org53}Institut f\"{u}r Kernphysik, Westf\"{a}lische Wilhelms-Universit\"{a}t M\"{u}nster, M\"{u}nster, Germany
\item \Idef{org54}Institut Pluridisciplinaire Hubert Curien (IPHC), Universit\'{e} de Strasbourg, CNRS-IN2P3, Strasbourg, France
\item \Idef{org55}Institute for Nuclear Research, Academy of Sciences, Moscow, Russia
\item \Idef{org56}Institute for Subatomic Physics of Utrecht University, Utrecht, Netherlands
\item \Idef{org57}Institute for Theoretical and Experimental Physics, Moscow, Russia
\item \Idef{org58}Institute of Experimental Physics, Slovak Academy of Sciences, Ko\v{s}ice, Slovakia
\item \Idef{org59}Institute of Physics, Academy of Sciences of the Czech Republic, Prague, Czech Republic
\item \Idef{org60}Institute of Physics, Bhubaneswar, India
\item \Idef{org61}Institute of Space Science (ISS), Bucharest, Romania
\item \Idef{org62}Instituto de Ciencias Nucleares, Universidad Nacional Aut\'{o}noma de M\'{e}xico, Mexico City, Mexico
\item \Idef{org63}Instituto de F\'{\i}sica, Universidad Nacional Aut\'{o}noma de M\'{e}xico, Mexico City, Mexico
\item \Idef{org64}iThemba LABS, National Research Foundation, Somerset West, South Africa
\item \Idef{org65}Joint Institute for Nuclear Research (JINR), Dubna, Russia
\item \Idef{org66}Konkuk University, Seoul, South Korea
\item \Idef{org67}Korea Institute of Science and Technology Information, Daejeon, South Korea
\item \Idef{org68}KTO Karatay University, Konya, Turkey
\item \Idef{org69}Laboratoire de Physique Corpusculaire (LPC), Clermont Universit\'{e}, Universit\'{e} Blaise Pascal, CNRS--IN2P3, Clermont-Ferrand, France
\item \Idef{org70}Laboratoire de Physique Subatomique et de Cosmologie, Universit\'{e} Grenoble-Alpes, CNRS-IN2P3, Grenoble, France
\item \Idef{org71}Laboratori Nazionali di Frascati, INFN, Frascati, Italy
\item \Idef{org72}Laboratori Nazionali di Legnaro, INFN, Legnaro, Italy
\item \Idef{org73}Lawrence Berkeley National Laboratory, Berkeley, California, United States
\item \Idef{org74}Lawrence Livermore National Laboratory, Livermore, California, United States
\item \Idef{org75}Moscow Engineering Physics Institute, Moscow, Russia
\item \Idef{org76}National Centre for Nuclear Studies, Warsaw, Poland
\item \Idef{org77}National Institute for Physics and Nuclear Engineering, Bucharest, Romania
\item \Idef{org78}National Institute of Science Education and Research, Bhubaneswar, India
\item \Idef{org79}Niels Bohr Institute, University of Copenhagen, Copenhagen, Denmark
\item \Idef{org80}Nikhef, National Institute for Subatomic Physics, Amsterdam, Netherlands
\item \Idef{org81}Nuclear Physics Group, STFC Daresbury Laboratory, Daresbury, United Kingdom
\item \Idef{org82}Nuclear Physics Institute, Academy of Sciences of the Czech Republic, \v{R}e\v{z} u Prahy, Czech Republic
\item \Idef{org83}Oak Ridge National Laboratory, Oak Ridge, Tennessee, United States
\item \Idef{org84}Petersburg Nuclear Physics Institute, Gatchina, Russia
\item \Idef{org85}Physics Department, Creighton University, Omaha, Nebraska, United States
\item \Idef{org86}Physics Department, Panjab University, Chandigarh, India
\item \Idef{org87}Physics Department, University of Athens, Athens, Greece
\item \Idef{org88}Physics Department, University of Cape Town, Cape Town, South Africa
\item \Idef{org89}Physics Department, University of Jammu, Jammu, India
\item \Idef{org90}Physics Department, University of Rajasthan, Jaipur, India
\item \Idef{org91}Physik Department, Technische Universit\"{a}t M\"{u}nchen, Munich, Germany
\item \Idef{org92}Physikalisches Institut, Ruprecht-Karls-Universit\"{a}t Heidelberg, Heidelberg, Germany
\item \Idef{org93}Politecnico di Torino, Turin, Italy
\item \Idef{org94}Purdue University, West Lafayette, Indiana, United States
\item \Idef{org95}Pusan National University, Pusan, South Korea
\item \Idef{org96}Research Division and ExtreMe Matter Institute EMMI, GSI Helmholtzzentrum f\"ur Schwerionenforschung, Darmstadt, Germany
\item \Idef{org97}Rudjer Bo\v{s}kovi\'{c} Institute, Zagreb, Croatia
\item \Idef{org98}Russian Federal Nuclear Center (VNIIEF), Sarov, Russia
\item \Idef{org99}Russian Research Centre Kurchatov Institute, Moscow, Russia
\item \Idef{org100}Saha Institute of Nuclear Physics, Kolkata, India
\item \Idef{org101}School of Physics and Astronomy, University of Birmingham, Birmingham, United Kingdom
\item \Idef{org102}Secci\'{o}n F\'{\i}sica, Departamento de Ciencias, Pontificia Universidad Cat\'{o}lica del Per\'{u}, Lima, Peru
\item \Idef{org103}Sezione INFN, Bari, Italy
\item \Idef{org104}Sezione INFN, Bologna, Italy
\item \Idef{org105}Sezione INFN, Cagliari, Italy
\item \Idef{org106}Sezione INFN, Catania, Italy
\item \Idef{org107}Sezione INFN, Padova, Italy
\item \Idef{org108}Sezione INFN, Rome, Italy
\item \Idef{org109}Sezione INFN, Trieste, Italy
\item \Idef{org110}Sezione INFN, Turin, Italy
\item \Idef{org111}SSC IHEP of NRC Kurchatov institute, Protvino, Russia
\item \Idef{org112}SUBATECH, Ecole des Mines de Nantes, Universit\'{e} de Nantes, CNRS-IN2P3, Nantes, France
\item \Idef{org113}Suranaree University of Technology, Nakhon Ratchasima, Thailand
\item \Idef{org114}Technical University of Split FESB, Split, Croatia
\item \Idef{org115}The Henryk Niewodniczanski Institute of Nuclear Physics, Polish Academy of Sciences, Cracow, Poland
\item \Idef{org116}The University of Texas at Austin, Physics Department, Austin, Texas, USA
\item \Idef{org117}Universidad Aut\'{o}noma de Sinaloa, Culiac\'{a}n, Mexico
\item \Idef{org118}Universidade de S\~{a}o Paulo (USP), S\~{a}o Paulo, Brazil
\item \Idef{org119}Universidade Estadual de Campinas (UNICAMP), Campinas, Brazil
\item \Idef{org120}University of Houston, Houston, Texas, United States
\item \Idef{org121}University of Jyv\"{a}skyl\"{a}, Jyv\"{a}skyl\"{a}, Finland
\item \Idef{org122}University of Liverpool, Liverpool, United Kingdom
\item \Idef{org123}University of Tennessee, Knoxville, Tennessee, United States
\item \Idef{org124}University of the Witwatersrand, Johannesburg, South Africa
\item \Idef{org125}University of Tokyo, Tokyo, Japan
\item \Idef{org126}University of Tsukuba, Tsukuba, Japan
\item \Idef{org127}University of Zagreb, Zagreb, Croatia
\item \Idef{org128}Universit\'{e} de Lyon, Universit\'{e} Lyon 1, CNRS/IN2P3, IPN-Lyon, Villeurbanne, France
\item \Idef{org129}V.~Fock Institute for Physics, St. Petersburg State University, St. Petersburg, Russia
\item \Idef{org130}Variable Energy Cyclotron Centre, Kolkata, India
\item \Idef{org131}Vin\v{c}a Institute of Nuclear Sciences, Belgrade, Serbia
\item \Idef{org132}Warsaw University of Technology, Warsaw, Poland
\item \Idef{org133}Wayne State University, Detroit, Michigan, United States
\item \Idef{org134}Wigner Research Centre for Physics, Hungarian Academy of Sciences, Budapest, Hungary
\item \Idef{org135}Yale University, New Haven, Connecticut, United States
\item \Idef{org136}Yonsei University, Seoul, South Korea
\item \Idef{org137}Zentrum f\"{u}r Technologietransfer und Telekommunikation (ZTT), Fachhochschule Worms, Worms, Germany
\end{Authlist}
\endgroup

\end{document}